%% file: paper.tex
\input lanlmac
\input rgmac

\input epsf
\readdefs\writedefs


\ifx\href\undefined\def\href#1#2{{#2}}\fi
\def\spireshome{%
http://www.slac.stanford.edu/cgi-bin/spiface/find/hep/www?FORMAT=WWW&}
{\catcode`\%=12
\xdef\spiresjournal#1#2#3{\noexpand\href{\spireshome
                          rawcmd=find+journal+#1%2C+#2%2C+#3}}
\xdef\spireseprint#1#2{\noexpand\href{\spireshome rawcmd=find+eprint+#1%2F#2}}
\xdef\spiresreport#1{\noexpand\href{\spireshome rawcmd=find+rept+#1}}
}
\def\eprint#1#2{\spireseprint{#1}{#2}{#1/#2}}
\def\report#1{\spiresreport{#1}{#1}}
\edef\hashmark{\string#}
\def\secref#1{\href{\hashmark section.#1}{#1}}


\def\alphamsbar{\hbox{$\alpha_{\overline{MS}}$}}
\def\alphav{\hbox{$\alpha_v$}}
\def\MSbar{\hbox{$\overline{MS}$}}
\def\mbar{\hbox{$\overline{m}$}}
\def\lili{\hbox{$\{ U_i U_i \}$}}
\def\lilj{\hbox{$\{ U_i U_j \}$}}
\def\sli{\hbox{$ \{ S U_i   \}$}}
\def\ssli{\hbox{$ \{ SS, S U_i   \}$}}

\def\Ln{\mathop{\hbox{$ {\rm Ln}$}}}

\Title{
  LA UR-95-2355}{
Decay constants with Wilson fermions at $\beta=6.0$}

\centerline{
  Tanmoy Bhattacharya and Rajan Gupta}
\smallskip
\centerline{\it
  T-8, MS-B285, Los Alamos National Laboratory, Los Alamos, NM 87545}


\bigskip
\bigskip
\bigskip

We present results of a high statistics study of $f_\pi$, $f_K$,
$f_D$, $f_{D_s}$, and $f_V^{-1}$ in the quenched approximation using
Wilson fermions at $\beta=6.0$ on $32^3 \times 64$ lattices.  We find
that the various sources of systematic errors (due to setting the
quark masses, renormalization constant, and lattice scale) are now
larger than the statistical errors.  Our best estimates, without
extrapolation to the continuum limit, are $f_\pi=134(4)\ \MeV$, $f_K
=159(3)\ \MeV$, $f_D = 229(7)\ \MeV$, $f_{D_s} = 260(4)\ \MeV$, and
$f_V^{-1}(m_\rho) = 0.33(1)$, where only statistical errors have been
shown.  We discuss the extrapolation to the continuum limit by
combining our data with those from other collaborations.

\Date{30 NOV 1995.}
\baselineskip=12pt 

\newsec{Introduction}

Phenomenologically, $f_D,\ f_{D_s}, \ f_B$ and $f_{B_s}$ are essential
ingredients needed to determine the less well know elements of the
Cabibbo-Kobayashi-Maskawa mixing matrix.  As these heavy-light decay
constants are at best very poorly measured, there has been a large
effort by many lattice groups to estimate them from numerical
simulations.  Decay constants are amongst the most precise quantities
that one can calculate on the lattice and 
a recent review has been presented by Allton at LATTICE95 
\ref\allton{C.~Allton, \melbourne, \eprint{hep-lat}{9509084}.}.  In 
this paper we present results for $f_\pi,\ f_K,\ f_K/f_\pi, f_D,\
f_D/f_\pi,\ f_{D_s}$, $f_{D_s}/f_D$, and vector decay constant
$f_V^{-1}$ from simulations done on 170 $32^3 \times 64$ quenched
lattices at $\beta = 6.0$ using Wilson fermions. We emphasize that
extrapolations to the continuum limit, incorporating the results from
other collaborations, are not very reliable as the combined data do
not show an unambiguous pattern of $O(a)$ corrections.

Preliminary results from a
subset of 100 lattices were presented at the LATTICE94 meeting
\ref\TbRgBiel{T. Bhattacharya and R.~Gupta, 
\spireseprint{hep-lat}{9501016}{\bielefeld\ 935}.}. The raw lattice 
results have not changed significantly since then, however we now
present a more detailed analysis of the systematic errors.  We
estimate the uncertainty in the results due to extrapolation of the
lattice data to physical values of the quark masses, the
renormalization constants for the lattice currents, and the extraction
of the lattice scale. We find that these various systematic errors are
now much larger than the statistical errors.  Finite size errors, if
present, are smaller than the statistical errors.  Our
best estimates are now given in the scheme called $TAD1$ to evaluate
the renormalization constants for the axial and vector currents.

The details of the lattices and the calculation of the spectrum are
given in a companion paper \ref\rWHMus{T.~Bhattacharya, R.~Gupta,
G.~Kilcup, and S.~Sharpe, \report{LAUR-95-2354}.}. In section
\secref{2} we briefly summarize the lattice parameters, and in section
\secref{3} we describe the lattice methodology and the consistency
checks made to extract the decay constants using estimates from
different types of fits and interpolating operators.  The choice of
renormalization constants for the axial and vector currents, $Z_A$ and
$Z_V$, is discussed in section \secref{4}, the lattice scale in
section \secref{5}, and the quark masses in section \secref{6}.  In
section \secref{7} we compare the data with predictions of quenched
chiral perturbation theory. The extrapolation of the data to physical
values of quark masses is discussed in section \secref{8}, and our
best estimates at $\beta=6.0$ are summarized in section \secref{9}.
In section \secref{10} we compare our data with those from other
collaborations ( GF11 \ref\weindc{GF11 Collaboration,
\spiresjournal{Nuc.+Phys.}{B421}{217} {\NPB{421} (1994) 217}.}, JLQCD
\ref\rJLQCD{S. Hashimoto,\melbourne, 
\spireseprint{hep-lat}{9510033}{hep-lat/9510033}.}, and APE
\ref\rAPE{C. Allton, \etal, \dallas 456, and private communications.}. 
The MILC data presented in 
\ref\rMILCfb{C. Bernard \etal, \melbourne,
\spireseprint{hep-lat}{9509045}{hep-lat/9509045}.} are preliminary and 
therefore not included in this analysis.)  and extrapolate
the combined data to the continuum limit.  Finally, we present our
conclusions in section \secref{11}.

\newsec{LATTICE PARAMETERS}

The details of the 170 $32^3 \times 64$ gauge lattices used in this
analysis are given in \rWHMus.  We refer the interested reader to it
for further details of the signal in the 2-point correlation functions
and on the extraction of the spectrum.  In this paper we concentrate
on the analysis of systematic errors in decay constants associated
with fixing the quark masses $\mbar= (m_u+m_d)/2, \ m_s$ and $m_c$, the
renormalization constants $Z_A$ and $Z_V$, and the extrapolation to
physical masses and the continuum limit.

To calculate decay constants we used the Wuppertal source quark
propagators at five values of quark mass given by $\kappa = 0.135$
($C$), $0.153$ ($S$), $0.155$ ($U_1$), $0.1558$ ($U_2$), and $0.1563$
($U_3$).  These quarks correspond to pseudoscalar mesons of mass
$2835$, $983$, $690$, $545$ and $431$ $\MeV$ respectively where we
have used $1/a=2.33\GeV$ for the lattice scale.  We construct two
types of correlation functions, smeared-local ($\Gamma_{SL}$) and
smeared-smeared ($\Gamma_{SS}$) which are combined in different ways
to extract the decay constants as discussed below.  The three light
quarks allow us to extrapolate the data to the physical isospin
symmetric light quark mass $\mbar $, while the $C$ and $S$ $\kappa$
values are selected to be close to the physical charm and strange
quark masses.  The physical value of strange quark lies between $S $
and $U_1$ and we use these two points to interpolate to it.  In most
cases we find that extrapolation to \mbar\ can be done using the six
combinations of light quarks $U_1 U_1, \ U_1 U_2, \ U_1 U_3,\ U_2
U_2,\ U_2 U_3,\ U_3 U_3$.  For brevity we will denote this combination
by \lilj\ and the three degenerate cases by \lili.

\newsec{Lattice Method for calculating $f_{PS}$ and $f_V$}

The lattice definition of the pseudo-scalar decay
constant $f_{PS}$, using the convention that the experimental
value is $f_\pi=131$~MeV, is~\ref
\hamberparisi{
	H. Hamber and G. Parisi, \spiresjournal{Phys.+Rev.}{D27}{208}
        {\PRD{27}, 208 (1983)}.
}
\eqn\defnfpi{
f_\pi = 
{Z_A \langle 0 | A_4^{\rm local} | \pi (\vec p )\rangle  \over {E_\pi(\vec p)}} \ ,
}
where $Z_A$ is the axial current renormalization constant connecting the 
lattice scheme to continuum \MSbar.  In order
to extract $f_\pi$ we study, in addition to the 2-point correlation
functions $\Gamma$, two kinds of ratios of correlators:
\eqn\decayratio{\eqalign{
R_1(t) &= {\Gamma_{SL}(t) \over { \Gamma_{SS}(t) }}
          \quad {\buildrel {\scriptstyle t\to \infty } \over \sim} \quad
          {\langle 0 | {A_4}^{\rm local} | \pi \rangle \over  
          {\langle 0 | {A_4}^{\rm smeared} | \pi \rangle }}		\cr
R_2(t) &= {\Gamma_{SL}(t) \Gamma_{SL}(t) \over { \Gamma_{SS}(t) }}
          \quad {\buildrel {\scriptstyle t\to \infty } \over \sim} \quad
          {|\langle 0 | A_4^{\rm local} | \pi \rangle |^2 \over
	  {2 M_\pi}} \ e^{-M_\pi t}.					\cr
}}
In the case of $R_1$ 
we have to extract $\langle 0 | {A_4}^{\rm smeared} | \pi \rangle$
separately from the $\Gamma_{SS}$ correlator.  For each of the two
ratios, $R_1$ and $R_2$, the smeared source $J$ used to create the
pion can be either $\pi$ or $A_4$.  This gives four ways of calculating
$f_\pi$, which we label as
$f_\pi^a $ (using ratio $R_1$ with $J=\pi$), 
$f_\pi^b $ (using ratio $R_1$ with $J=A_4$), 
$f_\pi^c $ (using ratio $R_2$ with $J=\pi$), and 
$f_\pi^d $ (using ratio $R_2$ with $J=A_4$).  
The fifth way, $f_\pi^e$, consists of combining the mass and amplitude of the
2-point correlation functions $\langle A_4 P \rangle_{LS}$ and
$\langle P P \rangle_{SS}$, and the sixth way, $f_\pi^f$, uses $\langle A_4
A_4 \rangle_{LS}$ and $\langle A_4 A_4 \rangle_{SS}$.  

The lattice results for mesons at $\vec p = 0$ for the different
combinations of quarks are given in Table~\nameuse\tfpizerop\ using
the renormalization scheme $Z_{TAD1}$ defined in
Table~\nameuse\tZschemes.  All Errors are estimated by a single
elimination Jackknife procedure.  The results from the six ways of
combining the two correlators are mutually consistent.  Since the
different methods use the same correlators, the data are highly
correlated; however, consistent results do indicate that fits have
been made to the lowest state in each of these correlators and
reassure us of the statistical quality of the data.

\table\tfpizerop{
\input t_fpizerop.tex }
{\vtop{\advance\hsize by -\parindent \noindent 
The data, in lattice units, for the pseudo-scalar decay constant
$f_{PS}$ for the six different ways of combining the $SL$ and $SS$
correlators described in the text. The renormalization scheme used to
generate this data is $TAD1$ as described in Table~\nameuse\tZschemes,
and the meson mass used in the analysis is taken to be the pole
mass.}}

The results for $f_\pi$ using correlators at non-zero momentum are
given in Table~\nameuse\tfpiatp. The data show that in almost all
cases the results are consistent within $2\sigma$.  The most
noticeable differences are in the $\vec p = (2,0,0)$ values for
lighter quarks.  The signal in these channels is not very good and it
is likely that in these cases there exists contamination from excited
states over the range of time-slices to which fits have been made. We
regard the overall consistency of the data as another successful check
of the lattice methodology.  Henceforth we shall restrict the analysis
to $\vec p = (0,0,0)$ case as it has the best signal.

\table\tfpiatp{
\input t_fpiatp.tex }
{\vtop{\advance\hsize by -\parindent \noindent 
The data, in lattice units, for the pseudo-scalar decay constant
$f_{PS}$, averaged over the six different ways of combining the $SL$
and $SS$ correlators, measured at different momenta. The
renormalization scheme is $TAD1$ as described in
Table~\nameuse\tZschemes, and the meson mass used in the analysis is
taken to be the pole mass.}}

The dimensionless vector decay constants are defined as
\eqn\evecdcdef{
Z_V \langle 0 | {V_\mu}^{\rm local} | V \rangle \ 
          = \ {\epsilon_\mu M_V^2 \over f_V}
}
where $V_\mu$ is the vector current and $| V \rangle$ is the lowest
$1^-$ state with mass $M_V$. The experimental quantities are related
to $f_V^{-1}$ by 
\eqn\efrhofv{\eqalign{
f_\rho^{-1}     \ &= \ \hphantom{-}{ 1 \over \sqrt2} f_V^{-1}(M_\rho) 
                     \qquad \ \ \ = \ \hphantom{-} 0.199(5) , \cr
f_\phi^{-1}     \ &= \ - { 1 \over 3} f_V^{-1}(M_\phi)\phantom{_{/\psi}}
                     \qquad \ \  = \  -0.078(1) , \cr
f_{J/\psi}^{-1} \ &= \ \hphantom{-} { 2 \over 3} f_V^{-1}(M_{J/\psi}) 
                     \qquad \ \  = \ \hphantom{-} 0.087(3) , \cr
}}
where the values are calculated using the rate $\Gamma(V \to e^+e^-)$
given in the PDB94 
\ref\rPDB{\href{http://www-pdg.lbl.gov/contents_rpp.html}
{Particle Data Book, \PRD{50} (1994) 1173}.}.
We extract the
relevant matrix element in the same two ways as described in
Eq.~\decayratio\ for $f_{PS}$.  To study discretization errors we
study three lattice transcriptions of the vector current (local,
extended, and conserved),
\eqn\evectorcurrents{\eqalign{
V_\mu^{L}(x) \ &= \ \bar\psi(x) \gamma_\mu \psi (x) , \cr
V_\mu^{E}(x) \ &= \ \bar\psi(x) \gamma_\mu U_\mu (x) \psi (x+\hat\mu) +
          \bar\psi(x+\hat\mu) \gamma_\mu U_\mu^\dagger (x) \psi (x) , \cr
V_\mu^{C}(x) \ &= \ \bar\psi(x) (\gamma_\mu-r) U_\mu (x) \psi (x+\hat\mu) +
   \bar\psi(x+\hat\mu) (\gamma_\mu + r)\ U_\mu^\dagger (x) \psi (x) . \cr
}}
where for degenerate quarks the last form is the conserved current.
In Tables~\nameuse\tfrhotypes\ and \nameuse\tfrhoschemes\ we show the
lattice data for the 15 mass combinations as a function of the
different methods/currents, and versus the renormalization schemes for
the local current. Overall, the data show that the two methods in
Eq.~\decayratio\ give consistent results for all three currents.  The
results from the local and extended vector currents also agree, while
those from the conserved current are $\approx 10\%$ smaller.  These
points will be discussed in more detail later.

\table\tfrhotypes{
\input t_frhotypes.tex }
{\vtop{\advance\hsize by -\parindent \noindent 
Lattice data for the vector decay constant $f_V^{-1}$ for the two
different ways of combining the $SL$ and $SS$ correlators, and for the
three different lattice vector currents described in the text. The
renormalization scheme in all cases is $TAD1$ as described in
Table~\nameuse\tZschemes, and the meson mass used in the analysis is 
taken to be the pole mass.}}

\table\tfrhoschemes{
\input t_frhoschemes.tex }
{\vtop{\advance\hsize by -\parindent \noindent 
Lattice data for the vector decay constant $f_V^{-1}$ as a function 
of the different renormalization schemes given in Table~\nameuse\tZschemes. 
The results are for the local current, and the meson mass used in the analysis is 
taken to be the pole mass.}}

In order to extract results that can be compared with experiments we
analyze the data in terms of the five sources of systematic errors 
discussed below.  

\newsec{\bf The renormalization constant $Z_A$ and $Z_V$}

Reliable calculations of decay constants depend on our ability to
calculate the renormalization constants, $Z_A$ and $Z_V$, linking the
lattice and continuum regularization schemes.  In our analysis we use
1-loop matching with the tadpole subtraction scheme of
Lepage-Mackenzie. An outline of the scheme, which includes picking a
good definition of the lattice $\alpha_s$ and the scale $q^*$ at which
to evaluate it is as follows. Lepage and Mackenzie show that
$\alpha_v$ (to be defined below) is a better expansion parameter than
the bare lattice coupling. To pick the value of $q^*$ we need to know
the ``mean'' momentum flow relevant to a given matrix element. Again
it has been pointed out by Lepage and Mackenzie that $q^*$, estimated
by calculating the mean momentum in the loop integrals, is dominated
by tadpole diagrams which are lattice artifacts.  If the tadpoles are
not removed then this scale is typically $\pi / a$. They have proposed
a meanfield improved version of the lattice theory which removes the
contribution of tadpoles. The effect of this is three-fold. One, it
typically changes $q^*$ to $1 / a$, $i.e.$ the matching scale becomes
more infrared if the tadpole diagram is removed; second the
renormalization of the quark field changes from $\sqrt{2 \kappa} \to
\sqrt{8\kappa_c}\sqrt{1 - 3 \kappa / 4 \kappa_c} $;
and finally the perturbative expression for $8\kappa_c$ is combined with 
the coefficient of $\alpha_v$ in the one loop matching relations to remove 
the tadpole contribution.

To get $\alphamsbar(q^*)$ we use the following Lepage-Mackenzie
scheme. The coupling $\alphav$ is defined at scale $q = 3.41/a$ to be 
\eqn\ealphav{
\alpha_v({3.41\over a}) \ \big(1 - (1.19 + 0.017n_f) \alpha_v \big) \ 
= \ - {3 \over 4 \pi} {\rm ln} 
({1\over3} {\rm Tr } Plaq) ,
}
which is related to $\alphamsbar$ at scale $q = 3.41/a$ by
\eqn\ealphav{
{1 \over \alphamsbar}  = {1 \over \alpha_v}  + 0.822 .
}
We then run \alphamsbar\ from $q$ to $q^*$ by integrating the 2-loop 
$\beta$-function.  To translate the results from $q^*$ to any other 
point one uses the standard continuum running.

At the lowest order there are two equally good tadpole factors, $U_0 =
plaquette^{1/4}$ or $8 \kappa_c$. To the accuracy of the meanfield
improvement one expects $8 \kappa_c U_0 = 1$. Deviations from this
relation ($\approx 10\%$ for the Wilson action at $\beta=6.0$) are a
measure of possible residual errors. Writing the tadpole factor as 
$1- X \alpha_{\MSbar}(q^*)$, we define a given Z factor to be
\eqn\eZdefn{
Z \ = \ 1 + \alpha_{\MSbar}(q^*) \bigg( {\gamma_0 \over 4\pi} \log(q^*a) + 
(C - X) \bigg) 
}
where $C$ is the difference between the finite part of
the continuum \MSbar\ and lattice 1-loop result.  Thus, $Z_A$ for 
the local operator in the tadpole improved schemes is 
\eqn\eZArenorm{
\sqrt{Z_\psi^1 Z_\psi^2} Z_A^L \ = \ 
                        \sqrt{1 - 3 \kappa_1 / 4 \kappa_c} 
		        \sqrt{1 - 3 \kappa_2 / 4 \kappa_c}  \ 
			\big( 1 -  \alpha_{\MSbar}(q^*) (1.68 - X) \big) .
}

In order to examine
the dependence of the decay constants on $Z$ and the renormalization
of the quark field we present our results for seven different commonly
used schemes described in Table~\nameuse\tZschemes.
The schemes $Z_{TADa}$, $Z_{TAD1}$, $Z_{TAD2}$, $Z_{TAD\pi}$, and
$Z_{Tgf11}$ are all self-consistent to $O(\alpha_s)$.  The scheme
$Z_{TADU_0}$ is ad hoc as we have replaced $8\kappa_c$ by $U_0$ in only
one part. We shall quote, as our best estimates, results obtained in
the $Z_{TAD1}$ scheme and use the difference between it and
$Z_{TAD\pi}$ as an estimate of the systematic error due to tuning
$q^*$.  Finally, an estimate of the residual perturbative errors is
taken to be the difference between $Z_{TAD1} $ and $Z_{TAD{U_0}}$, and
is given in column labeled $Z_A$ in Table~\nameuse\tdcfinal. This, we
believe, is an over-estimate of the error we make by using the 1-loop 
coefficient of $\alpha_v$. 

\table\tZschemes{
\input t_Zschemes.tex }
{\vtop{\advance\hsize by -\parindent \noindent %
The different renormalization schemes used in the analysis. The two
possible tadpole factors are $U_0 = plaq^{1/4} = 0.878$ and $1/8
\kappa_c = 0.795$. The 1-loop perturbative expansions for these are
$U_0 = 1 - 1.0492 \alpha$ and $8 \kappa_c = 1 + 1.364 \alpha$
respectively. The sixth scheme $Z_{Tgf11}$ is the one used by the 
GF11 collaboration with a slightly different definition of 
\alphamsbar \weindc.}}

The renormalization of the local vector current, $Z_V^L$ proceeds in
the same way as $Z_A$. In case of both the extended and conserved
currents, there is no tadpole contribution in $C$ as it 
cancels between the wave-function
renormalization and the vertex correction. Consequently, we use the
non-perturbative value for ${8\kappa_c}$. The complete renormalization
factors in the tadpole improved schemes for relating the lattice
results to the continuum are
\eqn\eZVrenorm{\eqalign{
\sqrt{Z_\psi^1 Z_\psi^2} Z_V^L \ &= \ 
                        \sqrt{1 - 3 \kappa_1 / 4 \kappa_c} 
		        \sqrt{1 - 3 \kappa_2 / 4 \kappa_c}  \ 
			\big( 1 -  \alpha_{\MSbar}(q^*) (2.182 - X) \big) , \cr
\sqrt{Z_\psi^1 Z_\psi^2} Z_V^E \ &= \ 
             {8\kappa_c}\sqrt{1 - 3 \kappa_1 / 4 \kappa_c} 
		        \sqrt{1 - 3 \kappa_2 / 4 \kappa_c}  \ 
			\big( 1 -  1.038 \alpha_{\MSbar}(q^*) \big) , \cr
\sqrt{Z_\psi^1 Z_\psi^2} Z_V^C \ &= \               
             {8\kappa_c}\sqrt{1 - 3 \kappa_1 / 4 \kappa_c} 
		        \sqrt{1 - 3 \kappa_2 / 4 \kappa_c} . \cr
}}
We find that the results with the local current lie in between those
from the extended and conserved currents, and have the best
statistical signal. We therefore quote results from the local current
as our best estimate, and use the difference between them as an
estimate of the systematic error.

\newsec{The lattice scale $a$}

To convert lattice results to physical units we use the lattice scale
extracted by setting $M_\rho$ to its physical value. This gives $1/a =
2.330(41)\ \GeV$ \rWHMus.  The variation of $1/a$ between the Jackknife
samples is folded into our error analysis, however different ways of
setting the scale are not. For example, using $M_N$ to
set the scale gives $1/a = 2.018(37)\ \GeV$ \rWHMus, while NRQCD simulations
of the charmonium and $\Upsilon$ spectrum give $1/a = 2.4(1)\ \GeV$
\ref\NRQCD{C.~T.~H. Davies, \etal, NRQCD collaboration, \spiresjournal{Phys.+Rev.}
{D50}{6963}{\PRD{50} (1994) 6963}.}. 
As we show later, the scale determined from $f_\pi$ is $2265(57)\
\MeV$.  Thus, estimates based on mesonic quantities like $M_\rho$,
heavy-heavy spectrum, and $f_\pi$ all give consistent results.  We 
take $1/a(M_\rho) = 2.330(41)\ \GeV$ and use
the spread, $\sim 70 \MeV \sim 3\%$, as our best guess of
the size of scaling violations relevant to the analysis of the decay
constants. To reduce this error requires using an improved gauge and
fermion action, which is beyond the scope of this work.

\newsec{Setting the quark masses}

In order to extrapolate the lattice data to physical values of the
quark mass we have to fix $\mbar$, $m_s$ and $m_c$. The chiral limit 
is determined by linearly extrapolating the data for 
$M_\pi^2$ to zero using the six cases \lilj. Our best estimate is 
\eqn\kappac{
\kappa_c   = 0.157131(9), 
}
which is used in the calculation of $Z_\psi$.

To fix the value of $\kappa_l$ corresponding to $\mbar$ we 
extrapolate the ratio $M_\pi^2/M_\rho^2$ to its physical value $0.03182$. 
The result is 
\eqn\kappal{
\kappa_{l} = 0.157046(9) 
}
Thus, our data is able to resolve between the chiral limit and
$\mbar$.  In \rWHMus\ we had shown that a non-perturbative estimate of
quark mass $m_{np}$, calculated using the Ward identity, and
$(1/2\kappa - 1/2\kappa_c)$ are linearly related for light quarks, so
either definition of the quark mass can be used for the extrapolation.
We have chosen to use $m_{np}$ in this paper. 

The determination of the strange quark mass has significant systematic
errors as shown below.  We determine $\kappa_s$ in three ways as
described in \rWHMus.  We extrapolate $M_K^2/M_\pi^2$, $M_{K^*} /
M_\rho$, and $M_{\phi} / M_\rho$ to $\mbar$ and then interpolate in
the strange quark to match their physical value. In
Table~\nameuse\tmstab\ we give $\kappa_s$, the non-perturbative
estimate $m_{np}a = m_s a$, and $m_{\rm s} = Z_m(1/2\kappa -
1/2\kappa_c)$ evaluated at $2\GeV$ in the $\overline{MS}$ scheme using
the $TAD1$ matching between lattice and continuum. The data show a
$\sim 20\%$ difference between various estimates of $m_s$ which cannot
be explained away as due to statistical errors.  Using $M_K^2/M_\pi^2$
to fix $m_s$ implies that $m_s \equiv 25 \mbar$ as we use the lowest
order chiral expansion to fit $M_{PS}^2$ data.  On the other hand
$M_{\phi} / M_\rho$ gives $m_s / \mbar \approx 30$.  This estimate is
not constrained by the chiral expansion, and is in surprisingly good
agreement with the next-to-leading chiral result
\ref\Donoghue{ J. Donoghue, B. Holstein, D. Wyler, 
\spiresjournal{Phys.+Rev.+Lett.}{69}{3444}{\PRL{69} (1992) 3444}.}. 
In this paper we shall quote results for both $m_{\rm s}(M_{K})$ and
$m_{\rm s}(M_{\phi})$, and take the values with $m_{\rm s}(M_{\phi})$
as our best estimates. The difference in results between using $m_{\rm
s}(M_{K})$ and $m_{\rm s}(M_{\phi})$ will be taken as an estimate of
the systematic error due to the uncertainty in setting $m_s$.

\table\tmstab{
\input t_mstab.tex }
{\vtop{\advance\hsize by -\parindent \noindent %
Estimates of $m_s$
using different combinations of hadron masses.  We give the $\kappa$
values, the quark mass determined by the Ward identity, and $m_{\rm s}
= Z_m(1/2\kappa - 1/2\kappa_c)$ evaluated at $2\GeV$ in the
$\overline{MS}$ scheme, and using the $TAD1$ tadpole subtraction procedure.}}

To determine the value of $\kappa$ corresponding to $m_c$ we match
$M_D,\ M_{D^*}$ and $M_{D_s}$ as these are obtained from the same
2-point correlation functions as used to determine the decay
constants.  Unfortunately, as shown in \rWHMus, the estimate of
charmonium and D meson masses measured from the rate of exponential
fall-off of the 2-point function (pole mass or $M_1$) and those from the kinetic
mass defined as $M_2 \equiv (\partial^2 E / \partial p^2
|_{p=0})^{-1}$ are significantly different.  We find that the data for
the heavy-heavy and heavy-light $q \bar q$ combinations are consistent
with the nearest-neighbor symmetric-difference relativistic dispersion
relation $\sinh^2(E/2) - \sin^2(p/2) = \sinh^2(M/2)$, in which case
$M_2$, as defined above, is given by $\sinh M$.  The results for $M_1$
and $M_2$ for the $D$ states are given in Table~\nameuse\tDmass\ for
$\kappa = 0.135$ (we have simulated only one heavy quark mass).  The
data show that the experimental results lie between $M_1$ and $M_2$
for each of the three states, and the difference between $M_1$ and
$M_2$ is large and statistically significant.  The size of this
systematic error and the uncertainty in setting the scale $1/a$, makes
it difficult to fix $\kappa_{charm}$. We simply assume that
$\kappa=0.135$ corresponds to $m_c$, and quote final results using
$M_2$.  As an estimate of systematic errors associated with not
tuning $m_c$ we take the difference in results between using $M_1$ and
$M_2$ since we do not have access to the rate of variation of decay
constants in the vicinity of $m_c$.

\table\tDmass{
\input t_Dmass.tex }
{\vtop{\advance\hsize by -\parindent \noindent %
A comparison of lattice estimates of $D$ meson masses with the 
experimental data.  We show results for $M_1$ and $M_2$ and for 
the two different ways of setting $m_s$ described in the text.}}

\newsec{Quenched approximation}

In the last couple of years it has been pointed out by Sharpe and collaborators
\ref\SRScpt{ S. Sharpe, \spiresjournal{Phys.+Rev.}{D41}{3233}
{\PRD{41} (1990) 3233}, \spiresjournal{Phys.+Rev.}{D46}{3146}
{\PRD{46} (1992) 3146} \semi 
J.~Labrenz, S. Sharpe, \spireseprint{hep-lat}{9312067}{\dallas, 335} \semi
S.~Sharpe, 1994 TASI lectures, \eprint{hep-ph}{9412243}} 
and by Bernard and Golterman
\ref\CbMgcpt{C.~Bernard and M.~Golterman, \spiresjournal{Phys.+Rev.}{D46}{853}
{\PRD{46} (1992) 853} \semi
M.~Golterman,
\eprint{hep-lat}{9405002} and
\eprint{hep-lat}{9411005}.}\
that there exist extra chiral logs due to the $\eta'$, which is also a
Goldstone boson in the quenched approximation.  These make the chiral
limit of quenched quantities sick. To analyze the effects of
quenching Bernard and Golterman \CbMgcpt\ have constructed the ratio
\eqn\eBGrfpiQ{
R \equiv {f_{12}^2 \over f_{11'}f_{22'}}
}
applicable in a 4-flavor theory where $m_1 = m_{1'}$ and $m_2 =
m_{2'}$.  The advantage of this ratio in comparing full and quenched
theories is that it is free of ambiguities due to the cutoff $\Lambda$
in loop integrals, and $O(p^4)$ terms in the chiral Lagrangian.  
The chiral expression for $R$ in the quenched theory is
\eqn\eRquenched{
R^{Q} = 1 + \delta\ \bigg[ { m_{12}^2 \over (m_{11'}^2 - m_{22'}^2)}
                          \log  {m_{11'}^2 \over m_{22'}^2}  - 1 \bigg] + 
	             O((m_1 - m_2)^2) ,
}
where $\delta \equiv {m_0^2 / 24 \pi^2 f_\pi^2}$ parameterizes the
effects of the $\eta'$.  The analogous expression in full QCD is 
\eqn\eRfull{
R^{F} =    1 + { 1 \over 32 \pi^2 f^2}  \bigg[ 
           m_{11'}^2 \Ln {m_{11'}^2 \over m_{12}^2}  + 
           m_{22'}^2 \Ln {m_{22'}^2 \over m_{12}^2}  \bigg] +
	             O((m_1 - m_2)^2) .
}
The leading analytic corrections in both cases are $O((m_1 - m_2)^2)$ 
\ref\rSharpeR{S. Sharpe, private communications.}, and were not 
included in the analysis presented at LATTICE 94
\ref\chiralrg{R.~Gupta, \spireseprint{hep-lat}{9412078}{\NPBPS{42} (1995) 85}.}.
The data, shown in Figs.~\nameuse\fRquen\ and \nameuse\fRfull,
indicates the need for including them in the fits. ($X_{quenched}$ is
the coefficient of $\delta$ in Eq.~\eRquenched, and $X_{full}$ is the
complete chiral logarithm term in Eq.~\eRfull.) The fit to the
quenched expression, Fig.~\nameuse\fRquen, gives $\delta = 0.14(4)$.
The fit to full QCD expression has smaller $\chi^2$ if we leave the
intercept as a free parameter. In that case the fit gives $1.69(45)$
and not unity as required by Eq.~\eRfull.  Thus, the effect of chiral
logs is small, barely discernible from the statistical errors, and
partly due to normal higher order terms in the chiral expansion. We
shall therefore neglect the effects of quenched chiral logs in this
study, and only discuss deviations of $f_{PS}$ from a behavior linear
in $m_q$ at the appropriate places.  

The second consequence of using the quenched approximation is that the
coefficients in the chiral expansion are different in the quenched and 
full theories.  This difference can be evaluated by comparing quenched
and full QCD data, which is beyond the scope of this work. Thus, we
cannot provide any realistic estimates of errors due to using the
quenched approximation.

\figure\fRquen{\epsfysize=5in\epsfbox{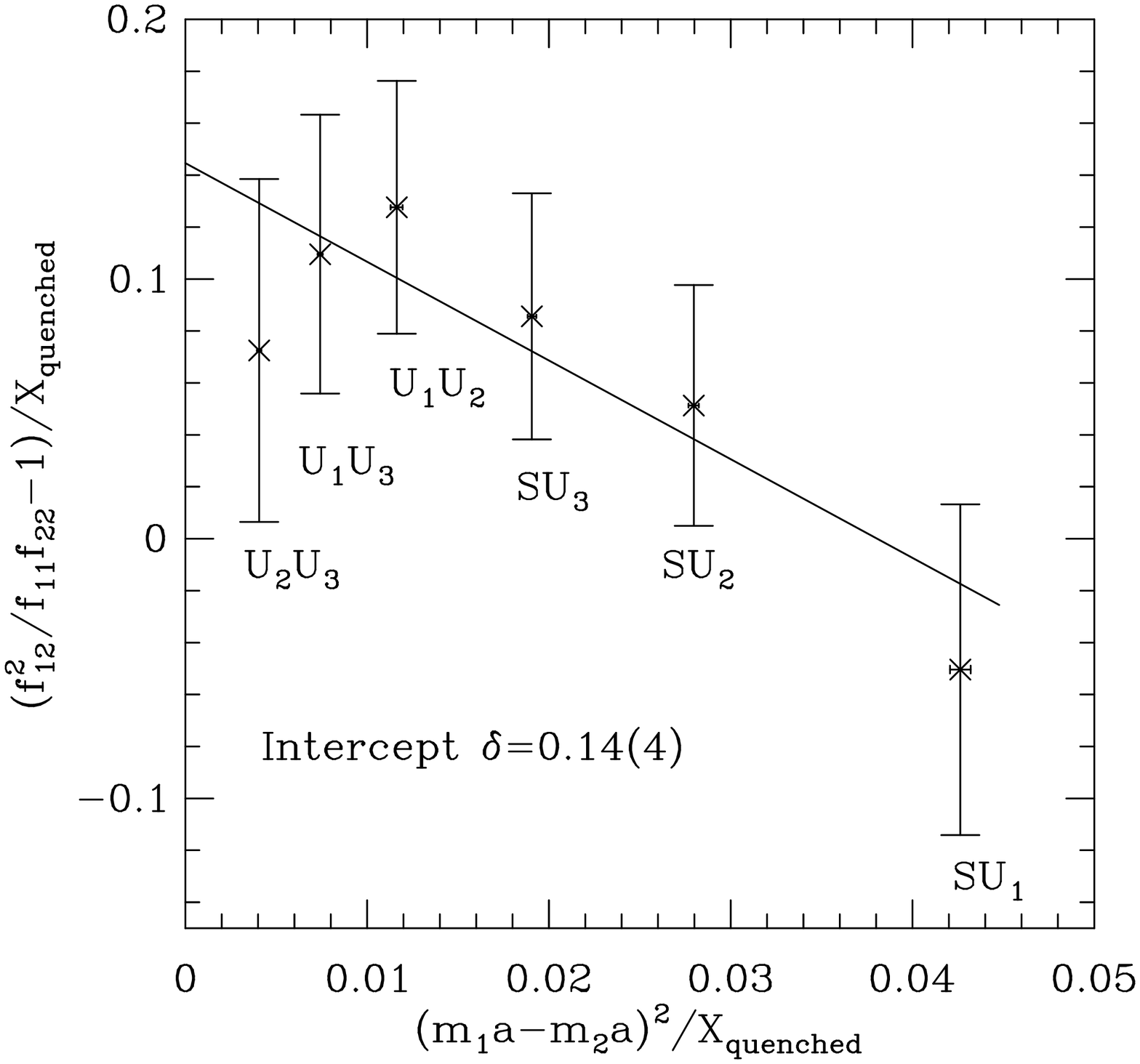}}
{\vtop{\advance\hsize by -\parindent \noindent 
Bernard-Golterman ratio $R$ versus $(m_1 - m_2)^2$. $X_{quenched}$ is
the coefficient of $\delta$ defined in Eq.~\eRquenched. The intercept 
gives $\delta = 0.14(4)$.}}

\figure\fRfull{\epsfysize=5in\epsfbox{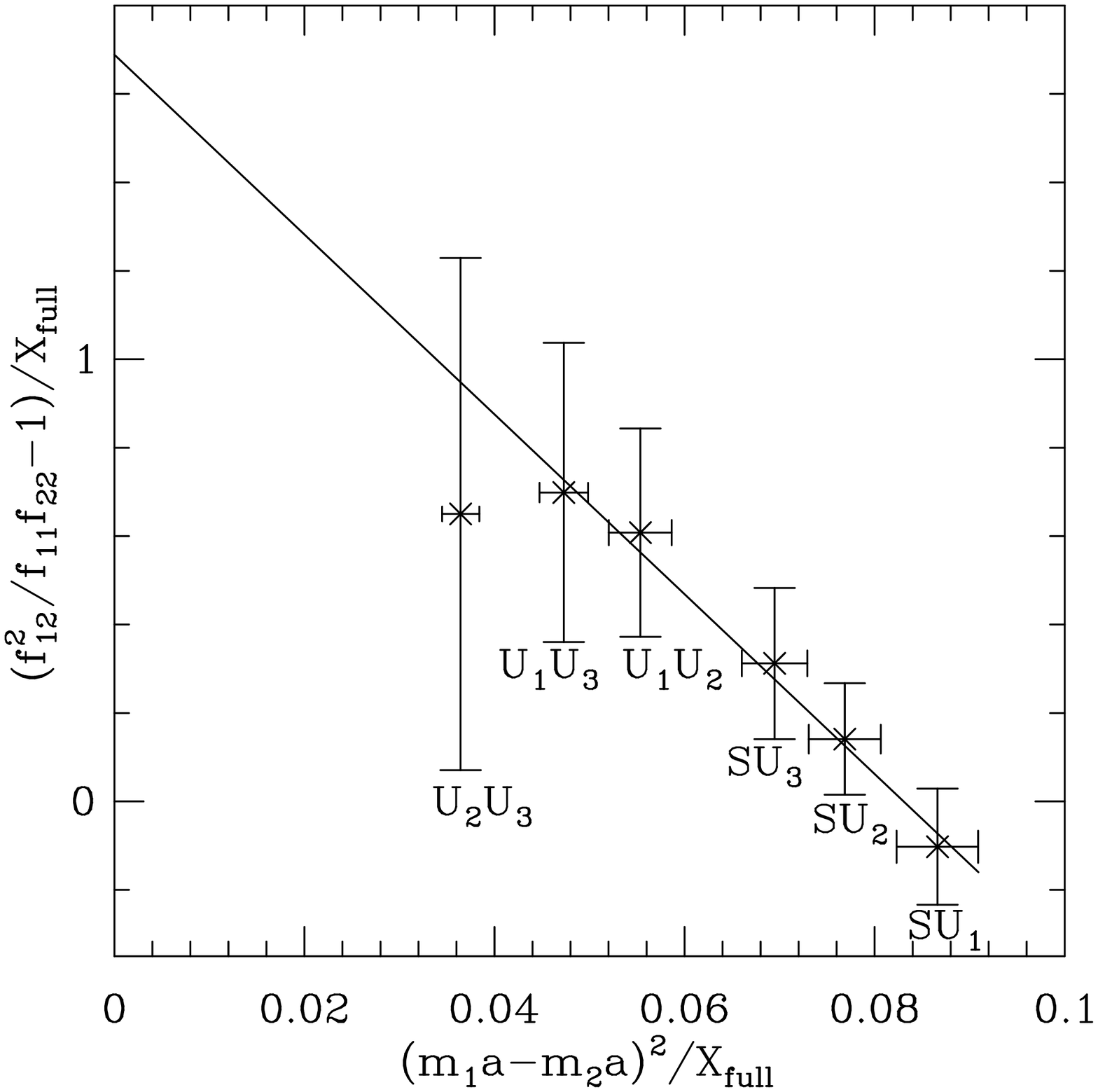}}
{\vtop{\advance\hsize by -\parindent \noindent 
Bernard-Golterman ratio $R$ versus $(m_1 - m_2)^2$. $X_{full}$ is the
chiral correction defined in Eq.~\eRquenched. The linear fit gives an
intercept of $1.69(45)$ instead of unity as indicated by
Eq.~\eRfull.}}

\figure\ffpiextrap{\epsfysize=5in\epsfbox{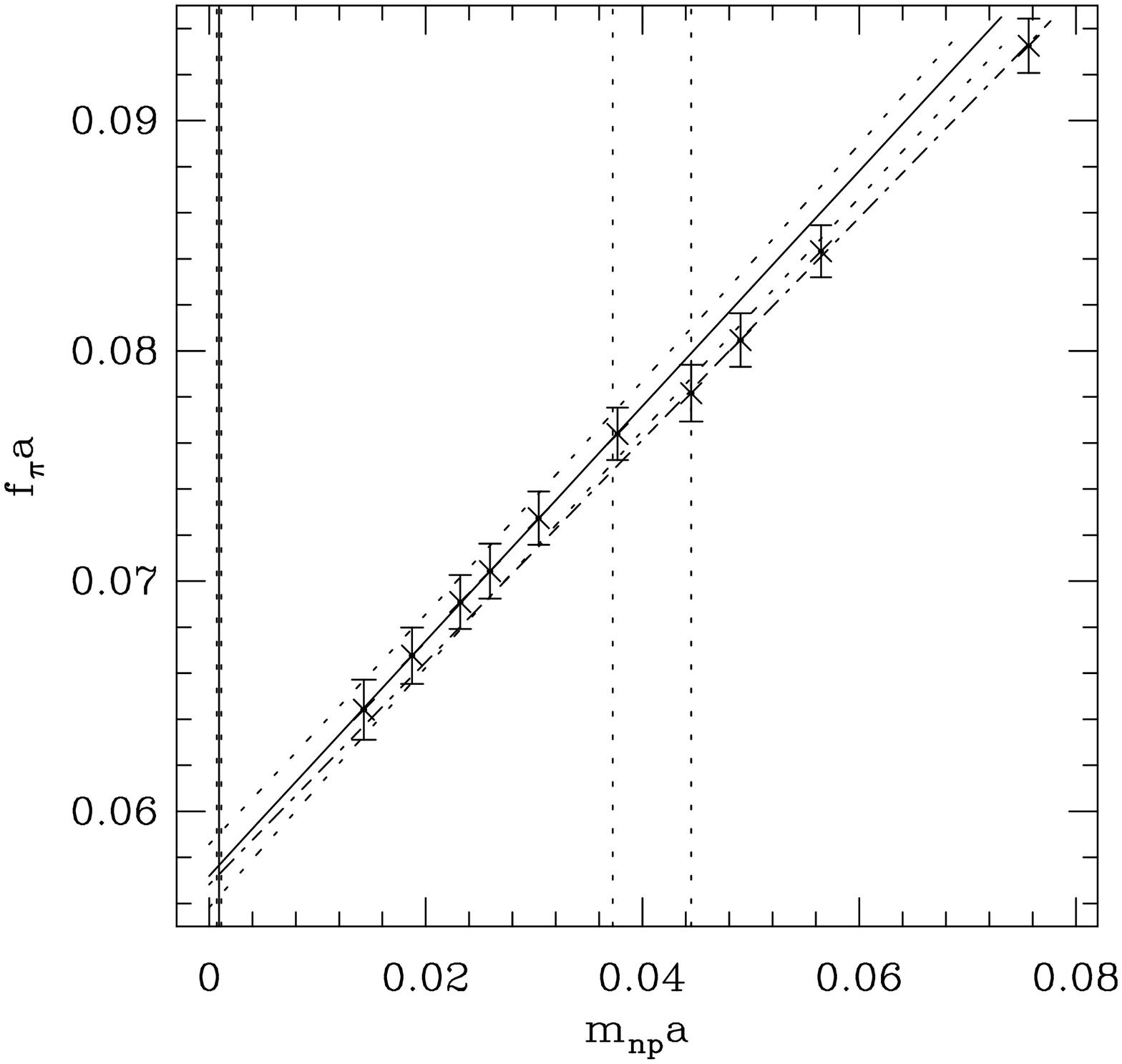}}
{\vtop{\advance\hsize by -\parindent \noindent %
Plot of 
data for $f_\pi a$ versus $m_{np}a$.  The linear fit, shown as a solid
line, is to the six \lilj\ points.  The error estimate on the fit is 
shown by the dotted lines. The dash-dot line is a linear fit to the four
$SS$ and \sli\ points. The vertical line at $m_{np}a \approx 0$ represents 
\mbar\ and the band at $m_{np}a \approx 0.04$ denotes the range of $m_s$.}}

\newsec{Extrapolation in quark masses}
In Fig.~\nameuse\ffpiextrap\ we show the pseudoscalar data for \lilj\ and \ssli\
combinations along with two different linear fits, one to the six
\lilj\ data points ($f_{PS} a= 0.0572(14) + 0.51(2) ma$) 
and the other to the four $SS$ and \sli\ points ($f_{PS}a = 0.0568(14)
+ 0.48(1) ma$).  Here $m$ is the average mass of the quark and anti-quark. 
The data show that even though the slopes for the two
fits are different, the values after extrapolation are virtually
indistinguishable.  The size of the break between the \ssli\ and
\lilj\ cases at $m_s$ is right at the $1\sigma$ level, and no such
break is visible between the $U_1U_i$ and the $U_2U_2$ cases.  We thus 
extrapolate to $f_\pi$ using \lilj\ points and assume that the overall
jackknife error adequately includes the uncertainty due to
extrapolation.

In Fig.~\nameuse\ffpsvslight\ we show the extrapolation for
heavy-light mesons for three cases of ``heavy'' ($C,\ S, \ U_1$)
quarks. The linear fits in the light quark mass, 
\eqn\efpsfits{\eqalign{
f_{PS} a&{}= 0.103(3) + 0.33(5) m_{np} a \qquad (CU_i) \,, \cr
f_{PS} a&{}= 0.074(1) + 0.26(1) m_{np} a \qquad (SU_i) \,, \cr
f_{PS} a&{}= 0.067(1) + 0.25(1) m_{np} a \qquad (U_1U_i) \,, \cr
}}
fit the data extremely well in each of the three cases.  Deviations from linearity
are apparent if the ``light'' quark mass is taken to be $S$ as shown
by the fourth point at $m_{np} a = 0.076$. These can be taken into
account by including corrections, $i.e.$ chiral logs and/or a
quadratic term. A fit including a quadratic term fits all four points
exceeding well, however the extrapolated value changes by $<0.2\sigma$
in all three cases.  Also, the change in curvature between $U_1U_i$
and $CU_i$ is within the error estimates. Considering that the form of
the correction term is not unique, and that the linear and quadratic
fits give essentially the same result, we consider it sufficient to
use a linear fit to the three $U_i$ points to extrapolate the
heavy-light decay constants to $\mbar$.

The difference in slope between fits to \lilj\ and \sli\ points does
effect the value of $f_K$. We, therefore, calculate it in two ways;
the central value is taken by extrapolating the \sli\ and $\{U_1U_i\}$
data in the light quark to \mbar\ and then interpolating in the
``heavy'' to $m_s$. In the second way we use the slope determined from
\lilj\ points and extrapolate to $\mbar+m_s$.  The two give consistent
results and we use the difference as an estimate of the systematic
error.

\figure\ffpsvslight{\epsfysize=5in\epsfbox{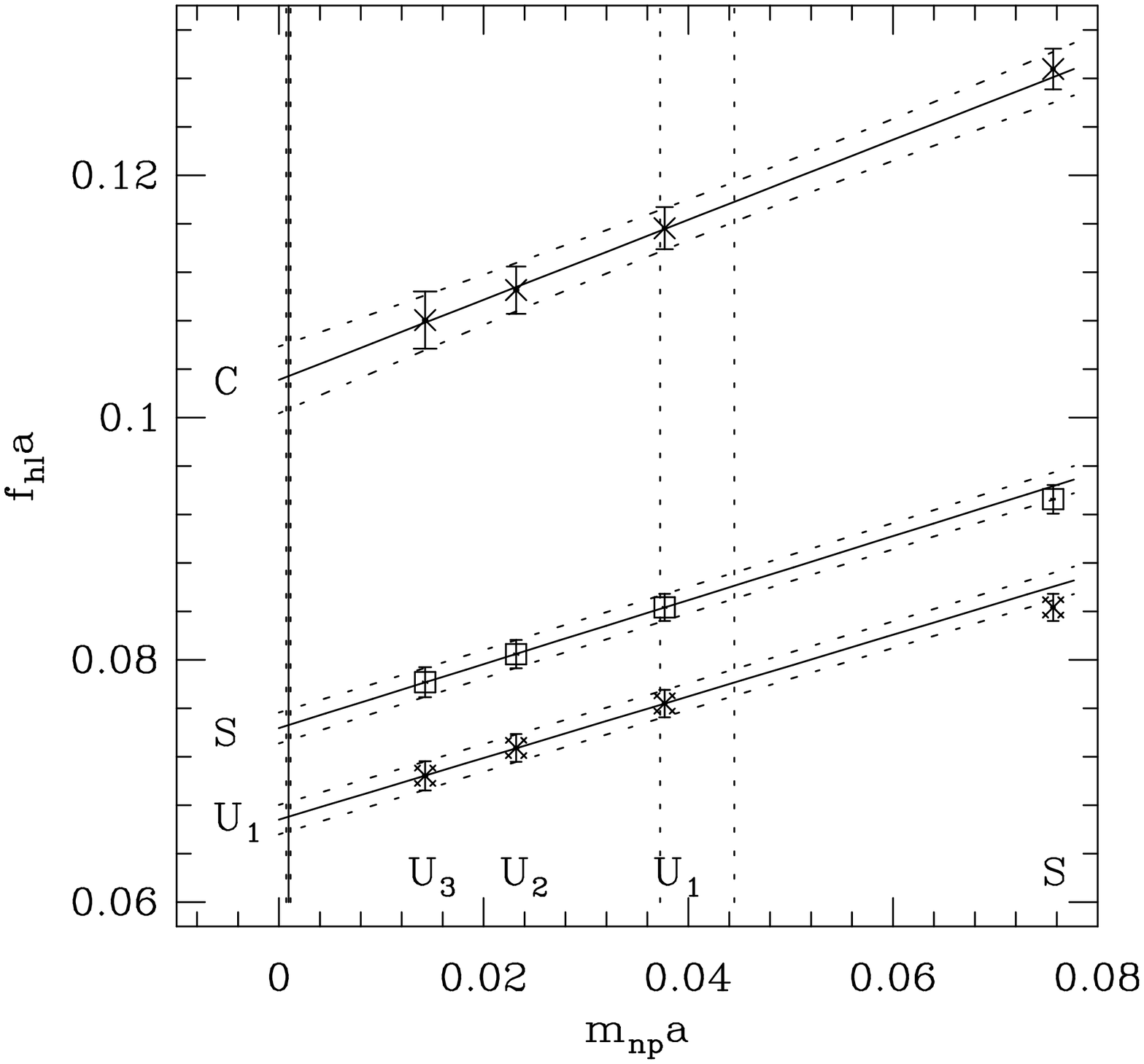}}
{\vtop{\advance\hsize by -\parindent \noindent %
Extrapolation of heavy-light pseudoscalar decay constants for 
three cases of ``heavy'', $C,\ S, \ U_1$ quarks. The linear fits are to 
to the three ``light'' $U_i$ quarks, and the fourth  point (light quark 
is $S$) is included 
to show the breakdown of the linear approximation.}}

The analogous plots for $f_V^{-1}$ are shown in
Figs.~\nameuse\ffrhoextrap\ and \nameuse\ffVvslight. To extract
$f_\rho^{-1}$ we make linear fits to the six \lilj\ and the three
\lili\ points.  As shown in Fig.~\nameuse\ffrhoextrap, these two fits
are almost identical ($f_{V} a = 0.328(10) + 0.33(23) m_{np} a$) and
neither of them fits the data very well. The \sli\ points show a very
significant break from the \lilj\ points, so to extract $f_{K^*}$,
$f_{D^*}$ we use the fits shown in Fig.~\nameuse\ffVvslight. As in the
case of $f_{PS}$, a linear fit to the three cases ($CU_i, SU_i, U_1U_i$) works well.
The fit parameters are
\eqn\efrhofits{\eqalign{
f_{V} a&{}= 0.163(6) + 0.31(10) \hphantom{0} m_{np} a  \qquad (CU_i)   \,, \cr
f_{V} a&{}= 0.300(9) + 0.026(14)             m_{np} a  \qquad (SU_i)   \,, \cr
f_{V} a&{}= 0.322(7) - 0.18(10) \hphantom{0} m_{np} a  \qquad (U_1U_i) \,, \cr
}}
Note that the slope changes sign between the $SU_i$ and $U_1 U_i$
cases. Since the points at $m_{np} a =0.076$ ($S$) show deviations from
the linear fits, we do not include this point in our analysis.

\figure\ffrhoextrap{\epsfysize=5in\epsfbox{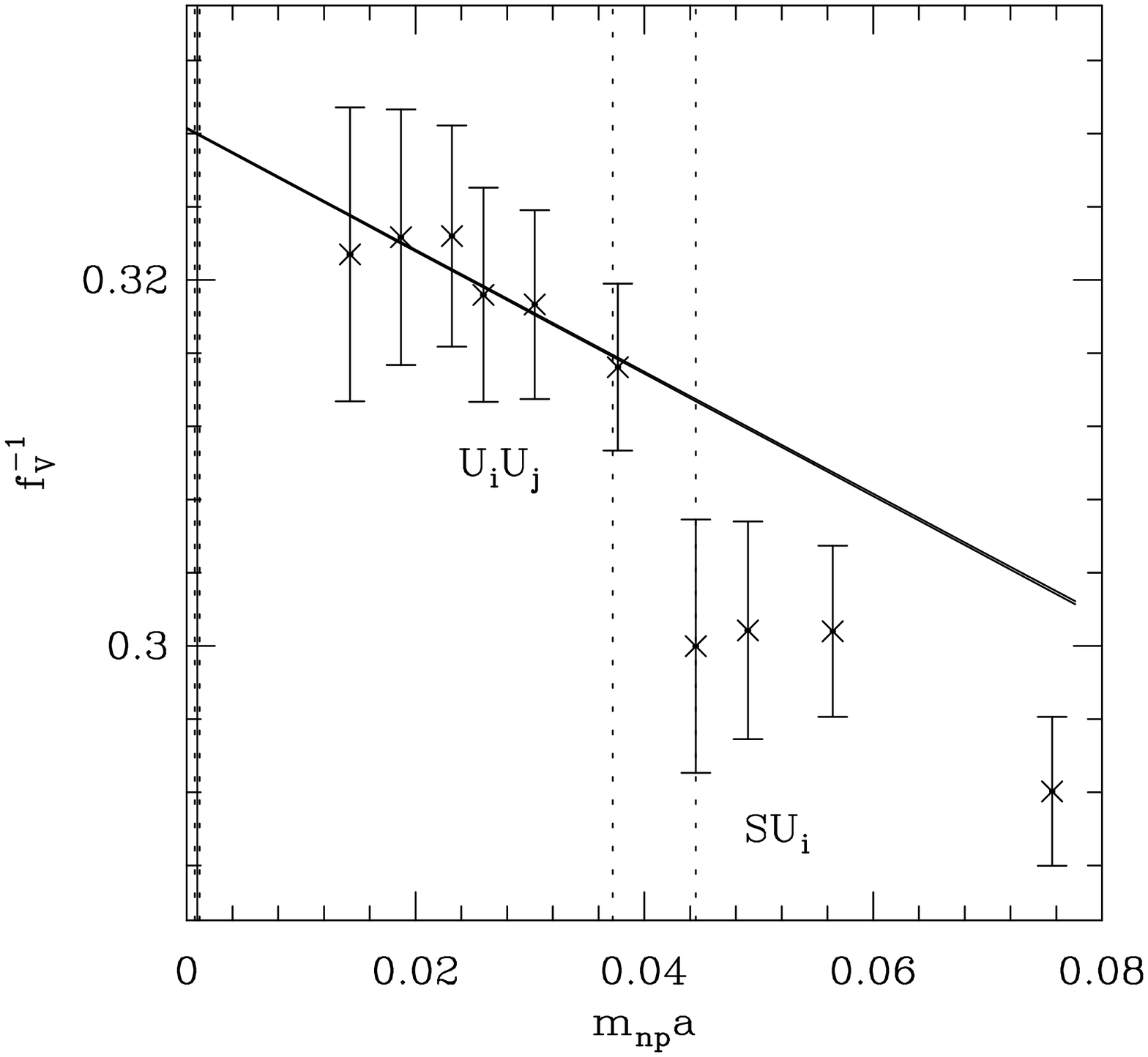}}
{\vtop{\advance\hsize by -\parindent \noindent 
Plot of data for $f_V^{-1}$ versus $m_{np} a$.  The linear fit is almost 
identical for the six \lilj\ or three \lili\ points. The $SS$ and \sli\ 
points are also shown for comparison.}}

\figure\ffVvslight{\epsfysize=5in\epsfbox{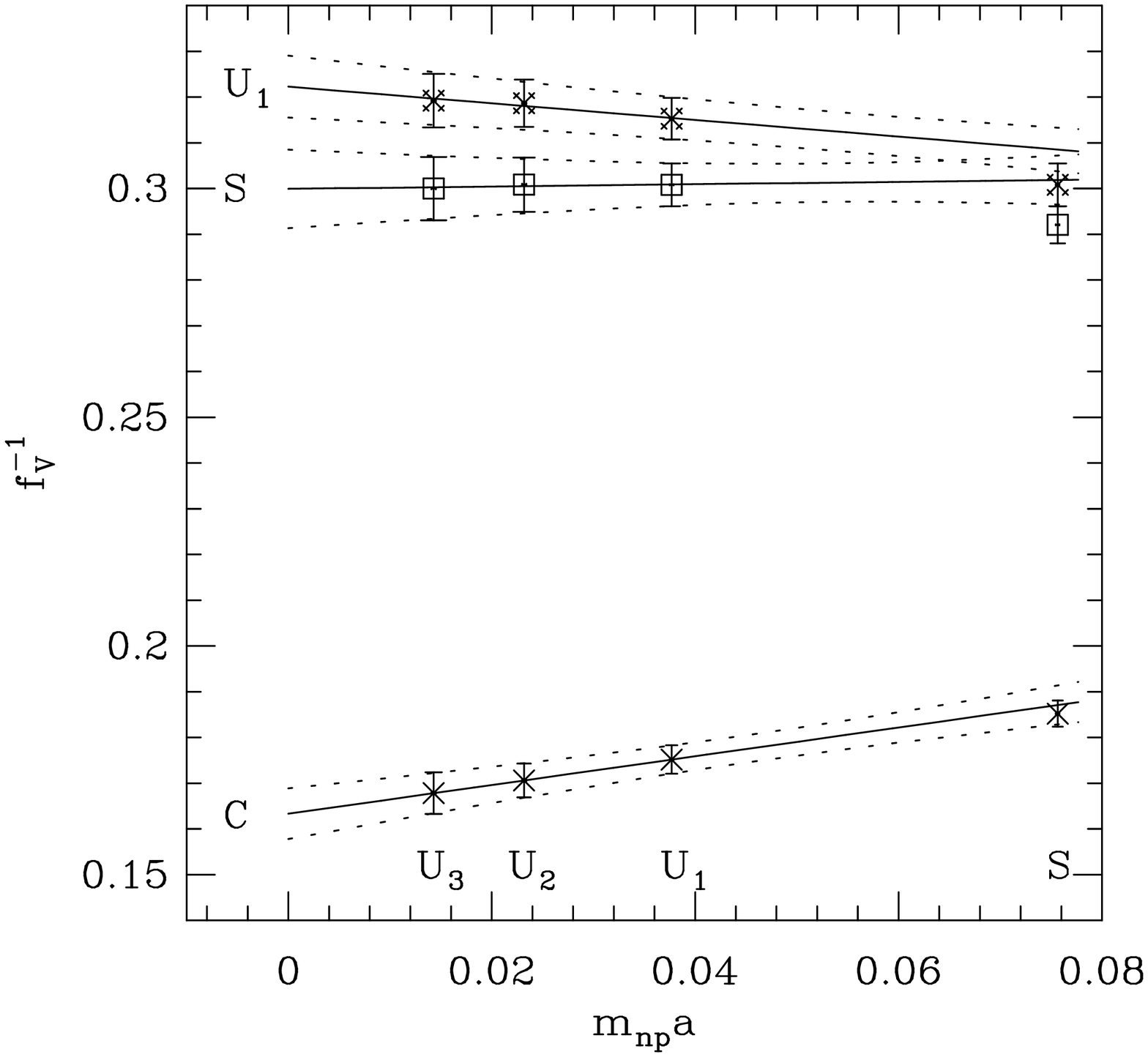}}
{\vtop{\advance\hsize by -\parindent \noindent %
Extrapolation of heavy-light vector decay constants for 
three cases of ``heavy'', $C,\ S, \ U_1$ quarks. The linear fits are to 
to the three ``light'' $U_i$ quarks, and the fourth  point (light quark 
is $S$) is included 
to show the breakdown of the linear approximation.}}

\newsec{Results at $\beta=6.0$}

The results for the pseudoscalar decay constants, in lattice units,
are given in Table~\nameuse\tfpilat\ for each of the seven
renormalization schemes.  The table also shows the variation with
respect to the two choices of $m_s$ and whether one uses $M_1$ or
$M_2$ for the heavy-light meson mass.  For our best estimates we use
$Z_{TAD1}$, and convert this data to $\MeV$ using $1/a(M_\rho)$.  The
results are summarized in Table~\nameuse\tdcTAD\ where we again
display variation with respect to $m_s$ and the heavy-light meson
mass.

\table\tfpilat{
\input t_fpilat.tex }
{\vtop{\advance\hsize by -\parindent \noindent 
Summary of results for pseudoscalar decay constants in
lattice units.  The variation with $m_s$ (set by $M_K$ or $M_\phi$)
and the heavy-light meson mass ($M_1$ or $M_2$) are shown explicitly.
The Jackknife error estimates include statistical and a part of systematic
errors due to extrapolation in quark masses.}}

Our final results are shown in Table~\nameuse\tdcfinal\ along with the
estimates of the various systematic errors discussed above. Thus, at 
$\beta = 6.0$ the value of $f_\pi$ come out about $3\%$ larger. Using
$f_\pi$ data to set the lattice scale gives $1/a(f_\pi) = 2265(57)\
\MeV$, whereas $1/a(M_\rho) = 2330(41)\ \MeV$ \rWHMus. 
Even ignoring the various systematic errors, the two estimates differ
by roughly $1\sigma$.

The ratio $f_K / f_\pi = 1.186(16)$ is about $2\sigma$ smaller than
the experimental value $1.223$ if one ignores all systematic errors.
(The systematic error in fixing $m_s$ would tend to lower our estimate,
$i.e.$ further increasing the difference.) An under-estimate of this
ratio in the quenched approximation is consistent with predictions of
quenched CPT \SRScpt\ \CbMgcpt.

\table\tdcTAD{
\input t_dcTAD.tex }
{\vtop{\advance\hsize by -\parindent \noindent 
Results for decay
constants in the $Z_{TAD1}$ scheme as a function of $m_s$ and
heavy-light meson masses. The data have been converted to $MeV$ using 
$M_\rho$ to set the scale. Only the jackknife error estimates are given.}}

\table\tdcfinal{
\input t_dcfinal.tex }
{\vtop{\advance\hsize by -\parindent \noindent 
Our final results using $TAD1$ scheme along with estimates of
statistical and various systematic errors as described in the text.
All dimensionful numbers are given in $MeV$ with the scale set by
$M_\rho$. For the systematic errors due to $m_s,\ m_c,\ q^*$ we also
give the sign of the effect.  We cannot estimate the uncertainty due
to using the quenched approximation.  Also, we do not have useful
estimates for entries marked with a ?.}}

The major uncertainty in the results for the heavy-light cases, $f_D$
and $f_{D_s}$, comes from the uncertainty in $Z_A$ and in setting the
charm mass. These corrections can be significant, and we 
need to reduce the various sources of systematic
errors in order to extract reliable continuum estimates.

In Tables ~\nameuse\tfVlat\ and \nameuse\tfVcurrent\ we give the
values for the vector decay constant $f_V^{-1}$, extrapolated to the
masses of a number of vector states even though some of them do not
decay electromagnetically to $l^+l^-$.  These tables also give the
variation with respect to setting $m_s$, the heavy-light meson mass
($M_1$ or $M_2$), $q^*$, $Z_A$, and the dependence on the lattice
current.  The criteria that the three types of currents should give
consistent results justifies using the Lepage-Mackenzie procedure for
$V_i^C$ also, as pointed out by Bernard in 
\ref\bernarddallas{C.~Bernard, \spireseprint{hep-lat}{9312086}{\dallas 47}.}. 
Using the $\sqrt{2\kappa}$ normalization for $V_i^C$
($i.e.$ the same normalization as $V_4^C(p_\mu=0)$ which is constrained by
the value of the conserved charge) gives significantly smaller values
for cases with $C$ quarks.

\table\tfVlat{
\input t_fVlat.tex }
{\vtop{\advance\hsize by -\parindent \noindent 
Results for $f_V^{-1}$ extrapolated to the masses of a number of
vector states specified within $[\ ]$  as a function of the renormalization 
schemes, $m_s$ ($M_K$ or $M_\phi$), and meson mass ($M_1$ or $M_2$).}}

\table\tfVcurrent{
\input t_fVcurrent.tex }
{\vtop{\advance\hsize by -\parindent \noindent 
Results for $f_V^{-1}$ as a function of the different discretizations 
of the vector current.  We also show the dependence on 
$m_s$ ($M_K$ or $M_\phi$) and meson mass ($M_1$ or $M_2$).}}

\newsec{Infinite volume continuum results}

In the companion paper analyzing the meson and baryon spectrum
\rWHMus\ we show that there are no noticeable differences between
results obtained on $24^3$ (earlier calculations) and our $32^3$
lattices. Thus we do not apply any finite size corrections to our
data.  To extract results valid in the 
continuum limit we combine our data with those from the GF11
($\beta=5.7, 5.93, 6.17$) \weindc, JLQCD ($\beta=6.1, 6.3$) \rJLQCD,
and APE ($\beta=6.0, 6.2$) \rAPE\ Collaborations. We have
attempted to correct for as many systematic differences, however some
like differences in lattice volumes, range of quark masses analyzed,
and fitting techniques remain.

We first compare the data for $f_\pi$ and $f_K$ from the different
collaborations as shown in Figs.~\nameuse\ffpiext\ and
\nameuse\ffkext.  The various calculations have similar statistics
(within a factor of two) and the two largest physical volumes used are
by GF11 ($24^3$ at $\beta=5.7$) and LANL ($32^3$ at $\beta=6.0$)
collaborations.  To facilitate comparison we make three changes; (a)
we switch to the convention in which $f_\pi = 93 \MeV $, (b) use the
$Z_{Tgf11}$ scheme, and (c) set $m_s$ using $M_K$.
%
%
A noticeable difference in the data shown in Figs.~\nameuse\ffpiext\
and \nameuse\ffkext\ is that the APE points at $\beta=6.0$ lie about
$1\sigma$ higher than LANL's and the value of $m_\rho a$ is also
larger. We believe that the difference is partly a result of
extrapolation from heavier quarks (APE collaboration use a linear fit
to extrapolate data at $\kappa=0.153, 0.154 0.155$ to the chiral
limit).  We find that both $f_{PS}$ and $M_V$ \rWHMus\ data show
negative curvature, and a linear extrapolation using only $SS$ and
$U_1U_1$ points increases LANL estimates, accounting for the full
difference in $M_\rho$ and a part of that in $f_\pi$. The more
important feature of the data, however is that neither plot shows a
clear $a$ dependence.  Nevertheless, a linear fit to all data,
assuming that lattice spacing errors are $O(a)$, gives
\eqn\efpifinal{\eqalign{
{f_\pi \over M_\rho} &= 0.110(5) \qquad \hbox{\rm (expt. 0.120)}, \cr
{f_K \over M_\rho}   &= 0.121(4) \qquad \hbox{\rm (expt. 0.147)}. \cr
}}
with $\chi^2/_{dof} = 1.6$ and $1.7$ respectively. The change from the
GF11 results is marginal as the fit is still strongly influenced by
the point at $\beta=5.7$, which may lie outside the domain of validity
of the linear extrapolation.  A linear extrapolation excluding the $\beta=5.7$
data gives 
\eqn\efpifinalB{\eqalign{
{f_\pi \over M_\rho} &= 0.118(10) , \cr
{f_K \over M_\rho}   &= 0.132(8)  , \cr
}}
with $\chi^2/_{dof} = 2.1$ and $1.9$ respectively.  
Using $m_s(M_\phi)$ would increase $f_K$ by $\sim 2\%$. 
Given the large
difference in the extrapolated value depending on whether the data at
$\beta=5.7$ is included or not makes it clear that more data are
required to make a reliable extrapolation to the continuum limit.

\figure\ffpiext{\epsfysize=5in\epsfbox{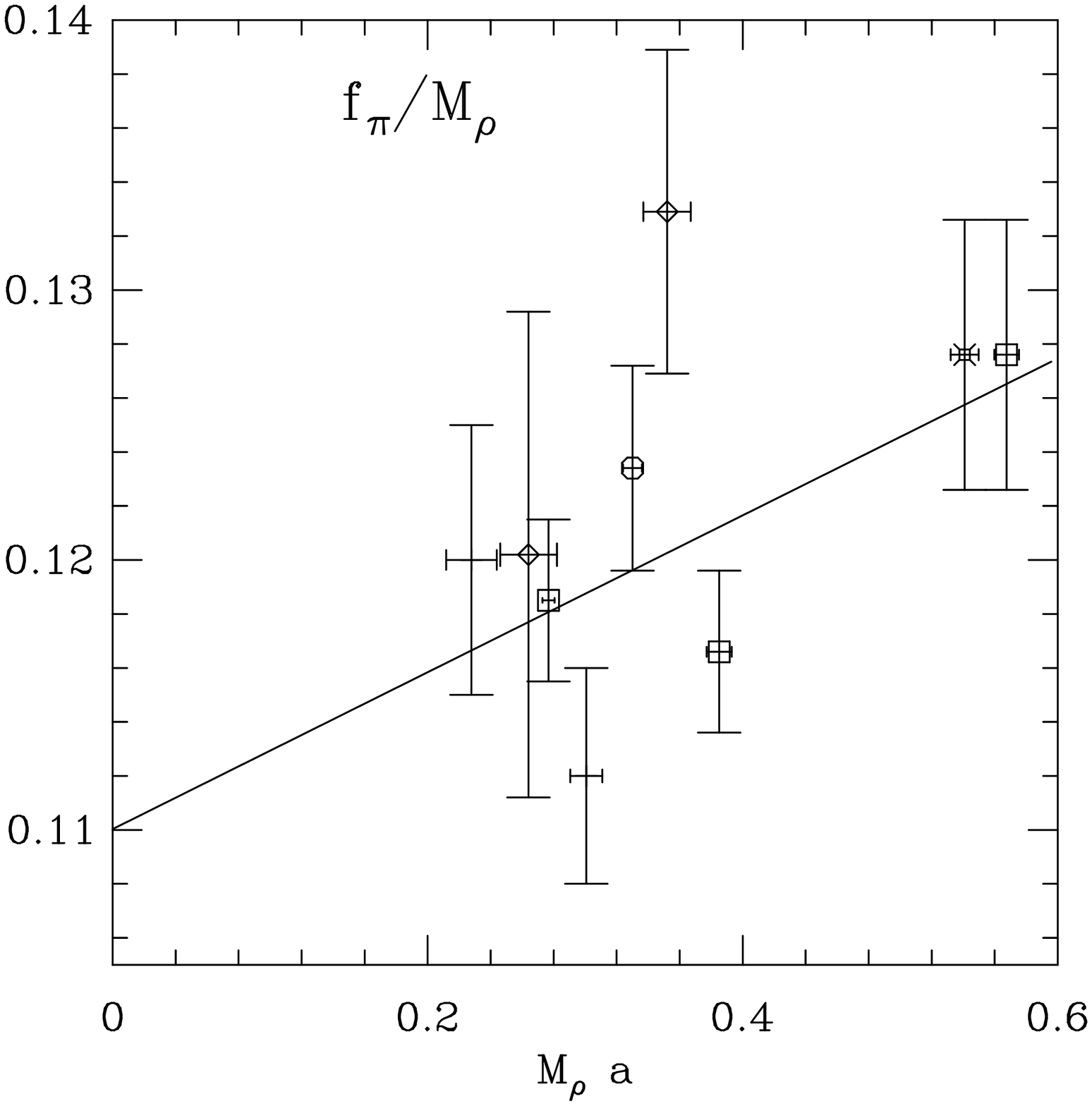}}
{\vtop{\advance\hsize by -\parindent \noindent %
Linear extrapolation to the continuum limit of the ratios 
$f_\pi / M_\rho $. Our data is shown with the symbol octagon, 
squares and fancy squares are the points from the GF11 Collaboration \weindc,
diamonds are APE collaboration points \rAPE, and the plus symbol labels 
JLQCD \rJLQCD\ data. The two GF11 points at $M_\rho a \approx 0.56$ represent
$16^3$ (squares) and $24^3$ (fancy squares) lattices at $\beta=5.7$.}}

\figure\ffkext{\epsfysize=5in\epsfbox{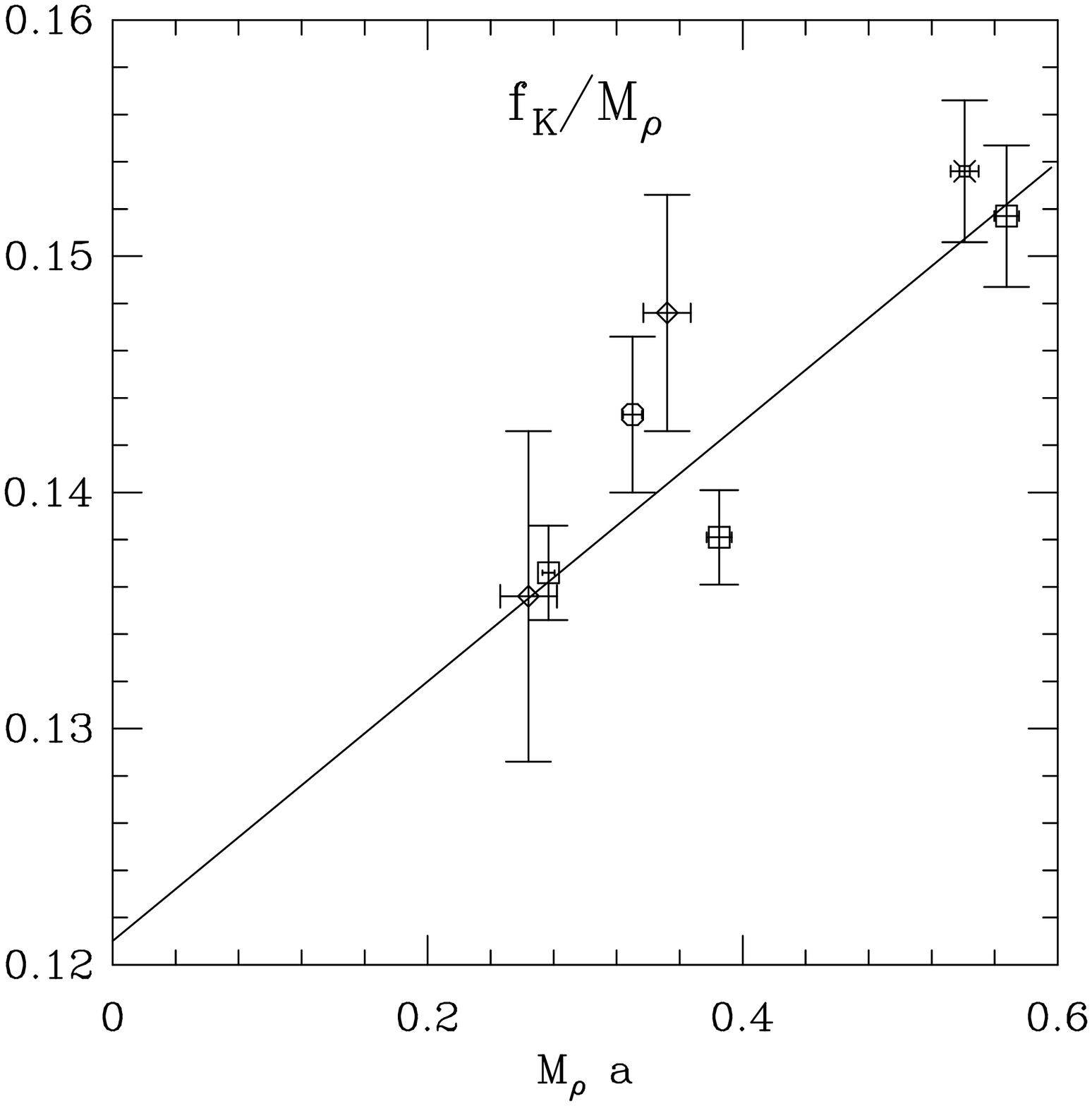}}
{\vtop{\advance\hsize by -\parindent \noindent %
Linear extrapolation to the continuum limit of the ratios 
$f_K / M_\rho $. Our data is shown with the symbol octagon, the 
squares and fancy squares are the points from the GF11 Collaboration \weindc, 
and the diamond labels APE \rAPE\ data.}}

The $f_D$ and $f_{D_s}$ data are combined with results from JLQCD
\rJLQCD\ and APE \rAPE\ collaborations as shown in
Fig.~\nameuse\fjlqcd.  The results are in ${TAD1}$ scheme, and for
comparison we use $m_s(M_K)$.  Also, from here on we switch back to
the convention in which $f_\pi=131 \MeV$. The APE collaboration use
$M_1$ for the meson mass. For consistency we have shifted their data
to $M_2$ using our estimates given in Table~\nameuse\tdcfinal.
%
%
A linear extrapolation to $a = 0$ then gives
\eqn\efDfinal{\eqalign{
f_D     &= 186(29) \MeV, \cr
f_{D_s} &= 218(15) \MeV, \cr
}}
with $\chi^2/_{dof} = 2.2$ and $2.0$ respectively. Using $m_s(M_\phi)$
would increase $f_{D_s}$ to $224(16)$ MeV. The quality of the fits
is, however, not very satisfactory. We feel that in order to improve the 
reliability of estimates in Eq.~\efDfinal\ one needs to 
reduce the various systematic errors that have not been included
in the $a\to 0$ extrapolations presented above. 

%

\figure\fjlqcd{\epsfysize=5in\epsfbox{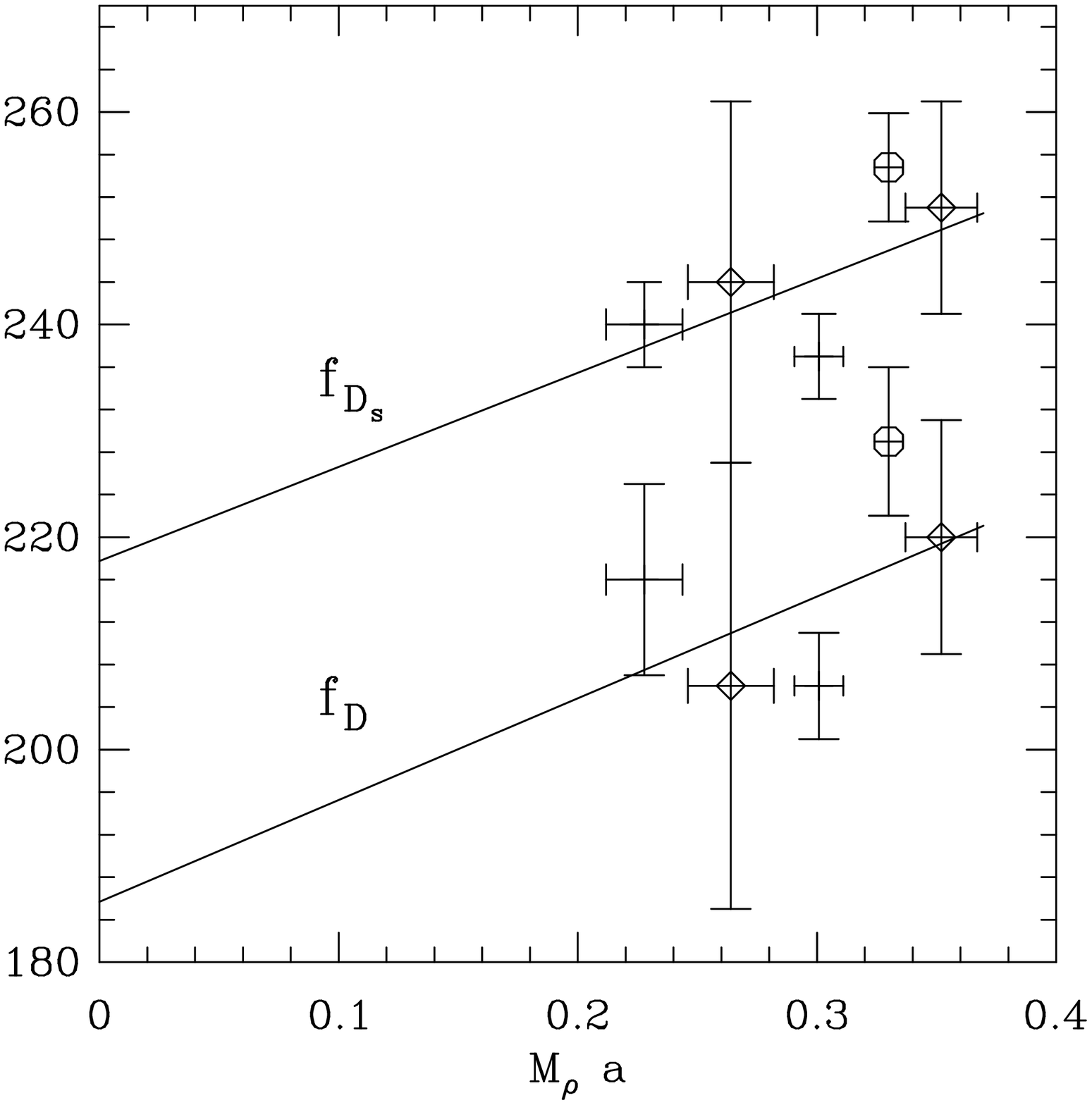}}
{\vtop{\advance\hsize by -\parindent \noindent %
Extrapolation to the continuum limit of $f_D $ and $f_{D_s}$ (in MeV) data. 
Our data is shown with the symbol octagon, 
the plus points are from the JLQCD Collaboration \rJLQCD,
and the diamonds label the APE collaboration \rAPE\ data.
}}

Finally, a linear fit to $f_\rho^{-1}$ data is shown in
Fig.~\nameuse\fweinV. The extrapolated value, $0.16(2)$ with
$\chi^2/_{dof} = 1.8$, is smaller than the experimental value
$0.199(5)$, and also smaller than that from a fit to just the GF11
data which gives $0.18(2)$ \weindc.

\figure\fweinV{\epsfysize=5in\epsfbox{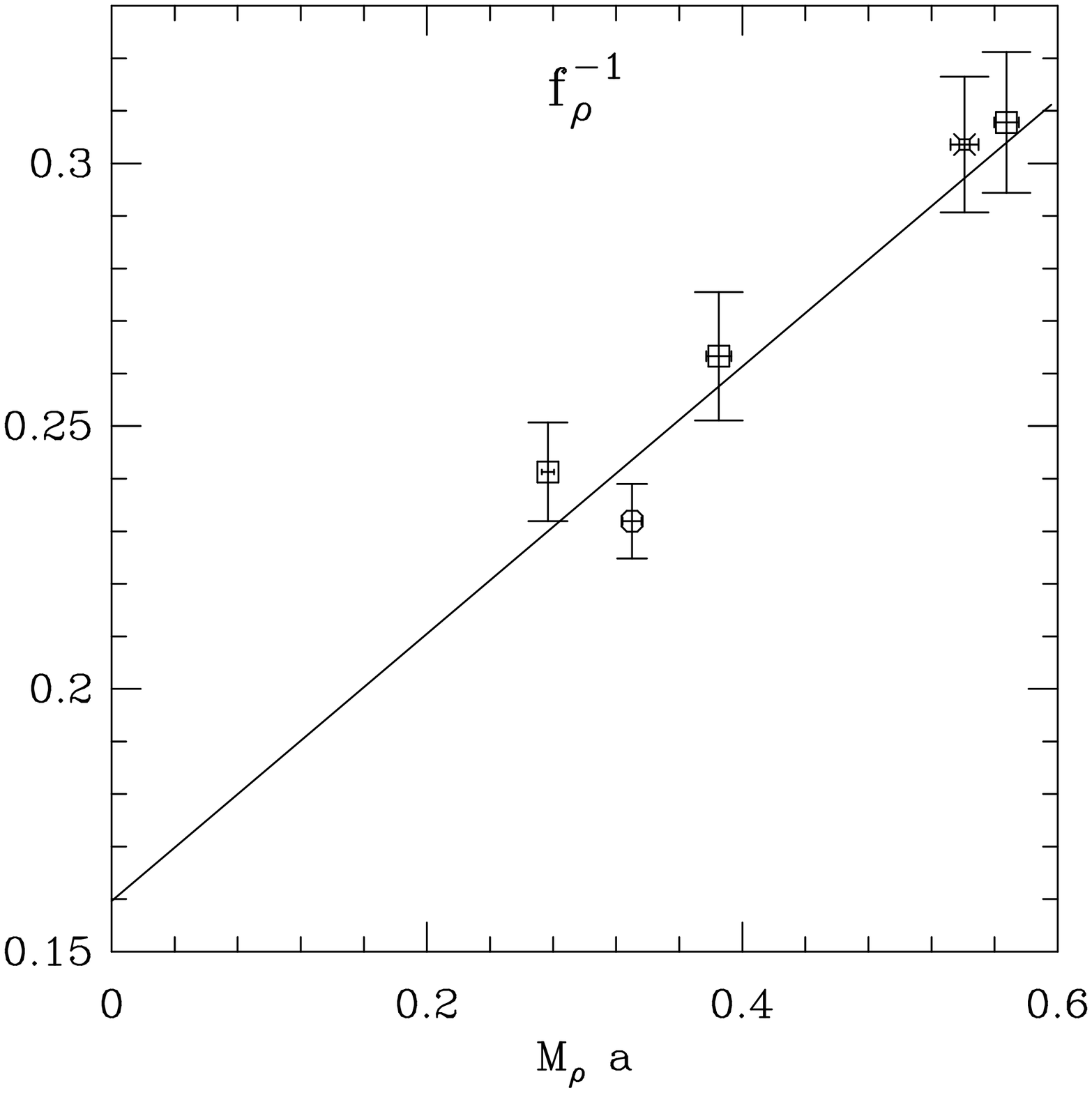}}
{\vtop{\advance\hsize by -\parindent \noindent %
Extrapolation to the continuum limit of $f_\rho^{-1}$. 
Our data is shown with the symbol octagon, 
and the rest of the points are from the GF11 Collaboration \weindc.}}

\newsec{Conclusions}

We have presented a detailed analysis of the decay constants involving
light-light and heavy-light (up to charm) quarks. We find that the
various sources of systematic errors (due to setting the quark masses,
renormalization constant, and lattice scale) are now larger than the
statistical errors.  Work is under progress to address these issues.
Our best estimates for the pseudo-scalar decay constants and the
various sources of error, without extrapolation to the continuum
limit, are given in Table~\nameuse\tdcfinal.

We would like to stress that including all of the present
high-statistics large lattice data, the extrapolation to the continuum
limit is, in all cases, not very reliable. For the Wilson action the
corrections are $O(a)$, and one expects that a linear extrapolation
should suffice starting at some $\beta$.  We find that in all cases
the combined world data do not show an unambiguous linear behavior in
$a$.  Since different groups analyze the data in different ways, 
there is no clean way of including the systematic errors in individual
points in the fits. We, therefore, cannot resolve whether the poor
quality of the linear fits is due to the various systematic and
statistical errors or due to the presence of higher order corrections.
As a result, our overall conclusion is that precise data at a few more
values of $\beta$ are required in order to extract reliable results in
the $a \to 0$ limit.

We have made linear fits to the data with and without including the
point at the strongest coupling, $\beta=5.7$.  A linear fit to
combined world data gives $f_\pi = 120(6)$ MeV and $f_K = 135(5)$ MeV.
Excluding $\beta=5.7$ point changes these estimates to $f_\pi =
128(6)$ and $f_K = 146(5)$ MeV.  Our best estimates for heavy-light
meson, $f_D= 186(29) \MeV$ and $f_{D_s} = 218(15) \MeV$ in the
continuum limit, are from a linear fit to data at $\beta \ge 6.0$. The
above estimates are using $m_s(M_K)$. Using $m_s(M_\phi)$ (our
preferred value) would increase $f_K$ and $f_{D_s}$ by $\approx 2\%$.

We study three lattice transcriptions of the vector current to calculate 
$f_V^{-1}$.  Using the Lepage-Mackenzie scheme to calculate $Z_V$ for 
each of the three currents yields results that are consistent to 
within $10\%$.  We extrapolate $f_\rho^{-1}$ to the continuum limit 
by combining with results from the GF11 collaboration. The 
result is $0.16(2)$ compared to the experimental value of $0.199(5)$.  

\newsec{Acknowledgements}

We are very grateful to Steve Sharpe and Claude Bernard for comments
on this paper, and to Chris Allton and Akira Ukawa for communicating
unpublished results of the APE and JLQCD collaborations to us. These
calculations have been done on the CM5 at LANL as part of the DOE HPCC
Grand Challenge program, and at NCSA under a Metacenter allocation.
We thank Jeff Mandula, Larry Smarr, Andy White and the entire staff at
the two centers for their tremendous support throughout this project.

\listrefs
\end

%% file: rgmac.tex

\def\GeV{\mathord{\rm \;GeV}}
\def\MeV{\mathord{\rm \;MeV}}

\def\sinh{\mathop{\rm sinh}\nolimits}
\def\bar{\overline}

\def\gsim{\mathrel{\raise2pt\hbox to 8pt{\raise -5pt\hbox{$\sim$}\hss{$>$}}}}
\def\rsim{\mathrel{\raise2pt\hbox to 8pt{\raise -5pt\hbox{$\sim$}\hss{$>$}}}}
\def\lsim{\mathrel{\raise2pt\hbox to 8pt{\raise -5pt\hbox{$\sim$}\hss{$<$}}}}
\def\ssqr#1#2{{\vbox{\hrule height.#2pt
      \hbox{\vrule width.#2pt height#1pt \kern#1pt\vrule width.#2pt}
      \hrule height.#2pt}\kern-.#2pt}}

\ifx\hyperref\undefined\def\hyperref#1#2#3#4{#4}\fi
\ifx\hyperdef\undefined\def\hyperdef#1#2#3#4{#4}\fi
\ifx\href\undefined\def\href#1#2{#2}\fi
\global\newcount\mtabno
\mtabno=1
\def\table#1#2#3{\DefWarn#1%
\xdef #1{\noexpand\hyperref{}{table}{\the\mtabno}%
{\the\mtabno}}\goodbreak\midinsert
$$#2$$\nobreak\centerline{Table~\hyperdef\hypernoname{table}%
{\the\mtabno}{\the\mtabno}. {\sl #3}}\bigskip\endinsert
\writedef{#1\leftbracket#1}\global\advance\mtabno by1}
\def\figure#1#2#3{\DefWarn#1\xdef#1{\noexpand\hyperref{}{figure}%
{\the\figno}{\the\figno}}\writedef{#1\leftbracket#1}%
\figinsert\figin{\centerline{#2}}\medskip\centerline{\vbox{\baselineskip12pt
\advance\hsize by -1truein\noindent\wrlabeL{#1=#1}\centerline{\sl
{\bf Fig.~\hyperdef\hypernoname{figure}{\the\figno}{\the\figno}:} #3}}}
\bigskip\endinsert\global\advance\figno by1}
\newread\myread
\def\ignore#1{}%
\def\readdefs{\ifx\writedef\ignore
 \immediate\openin\myread=\jobname.defs
 \ifeof\myread\message{No file \jobname.defs yet.}\else
   \closein\myread\relax\input\jobname.defs\fi\else
 \errmessage{can't \string\readdefs\space after \string\writedefs!!!}
\fi\relax}
\def\makelatexlike#1{\expandafter\let\csname old \string#1\endcsname=#1%
                 \def#1##1{%
                  \edef\tempname{\ifcat\relax\noexpand##1\noexpand##1\else
                                 \expandafter\noexpand\csname##1\endcsname\fi}%
                  \expandafter\expandafter\csname old \string#1\endcsname
                  \tempname}}
\def\nameuse#1{\edef\tempname{\ifcat\relax\noexpand#1\noexpand#1\else
                              \expandafter\noexpand\csname#1\endcsname\fi}%
               \expandafter\ifx\tempname\undefined
                  \message{YET UNDEFINED NAME \string`\string#1\string' USED.}%
                  ???%
               \else\expandafter\tempname\fi}








\def\author{\bigskip\centerline{David Daniel} 
    \smallskip\centerline{\it T-8, MS-B285,
                     Los Alamos National Laboratory, Los Alamos, NM 87545}}



\def\NPB#1{{\it Nucl. Phys.} {\bf B#1}}
\def\NPBPS#1{{\it Nucl. Phys.} {\bf B} ({\it Proc. Suppl.}) {\bf #1}}
\def\PRL#1{{\it Phys. Rev. Lett.} {\bf #1}}

\def\PRD#1{{\it Phys. Rev.} {\bf D#1}}

\def\etal{{\it et al.\ }}

\def\dallas{{\it ``LATTICE 93''}, Proceedings of
             the International Symposium on Lattice Field Theory, Dallas, 
             U.S.A., 1993, Eds. T.~Draper $et\ al.$, 
             \NPBPS{34}, (1994) }
\def\bielefeld{{\it ``LATTICE 94''}, Proceedings of
             the International Symposium on Lattice Field Theory, Bielefeld, 
             Germany, 1994, Eds. F.~Karsch $et\ al.$, 
             \NPBPS{42}, (1995) }

\def\melbourne{{\it ``LATTICE 95''}, To appear in Proceedings of
             the International Symposium on Lattice Field Theory, Melbourne,
             Australia, 1995, Eds. T.~D.~Kieu $et\ al.$, 
             \NPBPS{}, (1996)}



%% file: t_fpizerop.tex
\let\ifspace=\iffalse
\def\myskip{\omit&height1.5pt&%
\omit&&%
\omit&&%
\omit&&%
\omit&&%
\omit&&%
\omit&&%
 &\cr}
\vbox{\hbox{\vbox{
\tabskip=0pt\offinterlineskip
\def\tlr{\noalign{\hrule}}

\halign {\strut#& \vrule\vrule#\tabskip=3pt&
  \hfil$#$\hfil&\vrule#&
  \hfil$#$\hfil&\vrule#&
  \hfil$#$\hfil&\vrule#&
  \hfil$#$\hfil&\vrule#&
  \hfil$#$\hfil&\vrule#&
  \hfil$#$\hfil&\vrule#&
  \hfil$#$\hfil&\vrule#\tabskip=0pt\cr\tlr
\omit&height1.5pt&\multispan{13   }&\cr
\myskip
&& 
&& f_\pi^a                       
&& f_\pi^b                       
&& f_\pi^c                       
&& f_\pi^d                       
&& f_\pi^e                       
&& f_\pi^f                       
  &\cr
\myskip\tlr
\omit&height0.5pt&\multispan{13   }&\cr\tlr
\myskip
&&  ChCh                         
&& 0.198(2)
&& 0.198(3)
&& 0.197(2)
&& 0.197(2)
&& 0.197(3)
&& 0.197(3)
  &\cr\ifspace\myskip&&
&& 
&& 
&& 
&& 
&& 
&& 
  &\cr\fi\myskip\tlr
\myskip
&&  ChSt                         
&& 0.129(2)
&& 0.129(2)
&& 0.129(2)
&& 0.128(2)
&& 0.129(2)
&& 0.128(2)
  &\cr\ifspace\myskip&&
&& 
&& 
&& 
&& 
&& 
&& 
  &\cr\fi\myskip\tlr
\myskip
&&  ChU1                         
&& 0.115(2)
&& 0.116(2)
&& 0.116(2)
&& 0.115(2)
&& 0.116(2)
&& 0.115(2)
  &\cr\ifspace\myskip&&
&& 
&& 
&& 
&& 
&& 
&& 
  &\cr\fi\myskip\tlr
\myskip
&&  ChU2                         
&& 0.110(2)
&& 0.110(3)
&& 0.111(2)
&& 0.110(2)
&& 0.111(2)
&& 0.110(2)
  &\cr\ifspace\myskip&&
&& 
&& 
&& 
&& 
&& 
&& 
  &\cr\fi\myskip\tlr
\myskip
&&  ChU3                         
&& 0.107(3)
&& 0.108(3)
&& 0.109(2)
&& 0.108(2)
&& 0.109(2)
&& 0.108(2)
  &\cr\ifspace\myskip&&
&& 
&& 
&& 
&& 
&& 
&& 
  &\cr\fi\myskip\tlr
\myskip
&&  StSt                         
&& 0.093(1)
&& 0.093(1)
&& 0.093(1)
&& 0.093(1)
&& 0.094(2)
&& 0.093(2)
  &\cr\ifspace\myskip&&
&& 
&& 
&& 
&& 
&& 
&& 
  &\cr\fi\myskip\tlr
\myskip
&&  StU1                         
&& 0.084(1)
&& 0.084(1)
&& 0.085(1)
&& 0.084(1)
&& 0.085(2)
&& 0.084(1)
  &\cr\ifspace\myskip&&
&& 
&& 
&& 
&& 
&& 
&& 
  &\cr\fi\myskip\tlr
\myskip
&&  StU2                         
&& 0.081(1)
&& 0.080(1)
&& 0.081(1)
&& 0.081(1)
&& 0.081(2)
&& 0.080(2)
  &\cr\ifspace\myskip&&
&& 
&& 
&& 
&& 
&& 
&& 
  &\cr\fi\myskip\tlr
\myskip
&&  StU3                         
&& 0.078(1)
&& 0.078(1)
&& 0.079(1)
&& 0.078(1)
&& 0.078(2)
&& 0.077(2)
  &\cr\ifspace\myskip&&
&& 
&& 
&& 
&& 
&& 
&& 
  &\cr\fi\myskip\tlr
\myskip
&&  U1U1                         
&& 0.076(1)
&& 0.076(1)
&& 0.077(1)
&& 0.076(1)
&& 0.076(2)
&& 0.076(2)
  &\cr\ifspace\myskip&&
&& 
&& 
&& 
&& 
&& 
&& 
  &\cr\fi\myskip\tlr
\myskip
&&  U1U2                         
&& 0.073(1)
&& 0.072(1)
&& 0.073(1)
&& 0.073(1)
&& 0.073(2)
&& 0.072(2)
  &\cr\ifspace\myskip&&
&& 
&& 
&& 
&& 
&& 
&& 
  &\cr\fi\myskip\tlr
\myskip
&&  U1U3                         
&& 0.070(1)
&& 0.070(1)
&& 0.071(1)
&& 0.071(1)
&& 0.070(2)
&& 0.070(2)
  &\cr\ifspace\myskip&&
&& 
&& 
&& 
&& 
&& 
&& 
  &\cr\fi\myskip\tlr
\myskip
&&  U2U2                         
&& 0.069(1)
&& 0.069(1)
&& 0.070(1)
&& 0.069(1)
&& 0.069(2)
&& 0.069(2)
  &\cr\ifspace\myskip&&
&& 
&& 
&& 
&& 
&& 
&& 
  &\cr\fi\myskip\tlr
\myskip
&&  U2U3                         
&& 0.067(1)
&& 0.066(1)
&& 0.068(1)
&& 0.067(1)
&& 0.066(2)
&& 0.066(2)
  &\cr\ifspace\myskip&&
&& 
&& 
&& 
&& 
&& 
&& 
  &\cr\fi\myskip\tlr
\myskip
&&  U3U3                         
&& 0.064(1)
&& 0.064(1)
&& 0.066(1)
&& 0.065(1)
&& 0.064(3)
&& 0.064(2)
  &\cr\ifspace\myskip&&
&& 
&& 
&& 
&& 
&& 
&& 
  &\cr\fi\myskip\tlr
\cr}}}}
 

%% file: t_fpiatp.tex
\let\ifspace=\iffalse
\def\myskip{\omit&height1.5pt&%
\omit&&%
\omit&&%
\omit&&%
\omit&&%
\omit&&%
 &\cr}
\vbox{\hbox{\vbox{
\tabskip=0pt\offinterlineskip
\def\tlr{\noalign{\hrule}}

\halign {\strut#& \vrule\vrule#\tabskip=3pt&
  \hfil$#$\hfil&\vrule#&
  \hfil$#$\hfil&\vrule#&
  \hfil$#$\hfil&\vrule#&
  \hfil$#$\hfil&\vrule#&
  \hfil$#$\hfil&\vrule#&
  \hfil$#$\hfil&\vrule#\tabskip=0pt\cr\tlr
\omit&height1.5pt&\multispan{11   }&\cr
\myskip
&& 
&& {\ (\vec p = (0,0,0))}        
&& {\ (\vec p = (1,0,0))}        
&& {\ (\vec p = (1,1,0))}        
&& {\ (\vec p = (1,1,1))}        
&& {\ (\vec p = (2,0,0))}        
  &\cr
\myskip\tlr
\omit&height0.5pt&\multispan{11   }&\cr\tlr
\myskip
&&  ChCh                         
&& 0.198(2)
&& 0.198(2)
&& 0.203(2)
&& 0.200(3)
&& 0.201(3)
  &\cr\ifspace\myskip&&
&& 
&& 
&& 
&& 
&& 
  &\cr\fi\myskip\tlr
\myskip
&&  ChSt                         
&& 0.129(2)
&& 0.129(2)
&& 0.131(2)
&& 0.129(2)
&& 0.129(2)
  &\cr\ifspace\myskip&&
&& 
&& 
&& 
&& 
&& 
  &\cr\fi\myskip\tlr
\myskip
&&  ChU1                         
&& 0.116(2)
&& 0.116(2)
&& 0.118(2)
&& 0.115(2)
&& 0.115(2)
  &\cr\ifspace\myskip&&
&& 
&& 
&& 
&& 
&& 
  &\cr\fi\myskip\tlr
\myskip
&&  ChU2                         
&& 0.111(2)
&& 0.111(2)
&& 0.112(2)
&& 0.110(2)
&& 0.110(3)
  &\cr\ifspace\myskip&&
&& 
&& 
&& 
&& 
&& 
  &\cr\fi\myskip\tlr
\myskip
&&  ChU3                         
&& 0.108(2)
&& 0.108(3)
&& 0.110(3)
&& 0.107(3)
&& 0.108(3)
  &\cr\ifspace\myskip&&
&& 
&& 
&& 
&& 
&& 
  &\cr\fi\myskip\tlr
\myskip
&&  StSt                         
&& 0.093(1)
&& 0.094(1)
&& 0.095(2)
&& 0.096(2)
&& 0.099(2)
  &\cr\ifspace\myskip&&
&& 
&& 
&& 
&& 
&& 
  &\cr\fi\myskip\tlr
\myskip
&&  StU1                         
&& 0.084(1)
&& 0.085(1)
&& 0.086(2)
&& 0.087(2)
&& 0.090(2)
  &\cr\ifspace\myskip&&
&& 
&& 
&& 
&& 
&& 
  &\cr\fi\myskip\tlr
\myskip
&&  StU2                         
&& 0.080(1)
&& 0.081(2)
&& 0.082(2)
&& 0.083(2)
&& 0.086(2)
  &\cr\ifspace\myskip&&
&& 
&& 
&& 
&& 
&& 
  &\cr\fi\myskip\tlr
\myskip
&&  StU3                         
&& 0.078(1)
&& 0.079(2)
&& 0.080(3)
&& 0.080(2)
&& 0.083(2)
  &\cr\ifspace\myskip&&
&& 
&& 
&& 
&& 
&& 
  &\cr\fi\myskip\tlr
\myskip
&&  U1U1                         
&& 0.076(1)
&& 0.077(2)
&& 0.078(2)
&& 0.079(2)
&& 0.083(3)
  &\cr\ifspace\myskip&&
&& 
&& 
&& 
&& 
&& 
  &\cr\fi\myskip\tlr
\myskip
&&  U1U2                         
&& 0.073(1)
&& 0.073(2)
&& 0.074(2)
&& 0.075(3)
&& 0.080(3)
  &\cr\ifspace\myskip&&
&& 
&& 
&& 
&& 
&& 
  &\cr\fi\myskip\tlr
\myskip
&&  U1U3                         
&& 0.070(1)
&& 0.071(2)
&& 0.072(2)
&& 0.072(3)
&& 0.078(4)
  &\cr\ifspace\myskip&&
&& 
&& 
&& 
&& 
&& 
  &\cr\fi\myskip\tlr
\myskip
&&  U2U2                         
&& 0.069(1)
&& 0.070(2)
&& 0.071(2)
&& 0.071(3)
&& 0.077(4)
  &\cr\ifspace\myskip&&
&& 
&& 
&& 
&& 
&& 
  &\cr\fi\myskip\tlr
\myskip
&&  U2U3                         
&& 0.067(1)
&& 0.068(2)
&& 0.069(3)
&& 0.068(3)
&& 0.075(5)
  &\cr\ifspace\myskip&&
&& 
&& 
&& 
&& 
&& 
  &\cr\fi\myskip\tlr
\myskip
&&  U3U3                         
&& 0.064(1)
&& 0.066(2)
&& 0.067(3)
&& 0.064(4)
&& 0.074(5)
  &\cr\ifspace\myskip&&
&& 
&& 
&& 
&& 
&& 
  &\cr\fi\myskip\tlr
\cr}}}}
 

%% file: t_frhotypes.tex
$$
\let\ifspace=\iffalse
\def\myskip{\omit&height1.5pt&%
\omit&&%
\omit&&%
\omit&&%
\omit&&%
\omit&&%
\omit&&%
 &\cr}
\vbox{\hbox{\vbox{
\tabskip=0pt\offinterlineskip
\def\tlr{\noalign{\hrule}}

\halign {\strut#& \vrule\vrule#\tabskip=3pt&
  \hfil$#$\hfil&\vrule#&
  \hfil$#$\hfil&\vrule#&
  \hfil$#$\hfil&\vrule#&
  \hfil$#$\hfil&\vrule#&
  \hfil$#$\hfil&\vrule#&
  \hfil$#$\hfil&\vrule#&
  \hfil$#$\hfil&\vrule#\tabskip=0pt\cr\tlr
\omit&height1.5pt&\multispan{13   }&\cr
\myskip
&& 
&& f_\rho^{a}\ Loc.              
&& f_\rho^{b}\ Loc.              
&& f_\rho^{a}\ Ext.              
&& f_\rho^{b}\ Ext.              
&& f_\rho^{a}\ Con.              
&& f_\rho^{b}\ Con.              
  &\cr
\myskip\tlr
\omit&height0.5pt&\multispan{13   }&\cr\tlr
\myskip
&&  ChCh                         
&& 0.186( 02)
&& 0.186( 03)
&& 0.184( 03)
&& 0.185( 02)
&& 0.168( 03)
&& 0.169( 02)
  &\cr\ifspace\myskip&&
&& 
&& 
&& 
&& 
&& 
&& 
  &\cr\fi\myskip\tlr
\myskip
&&  ChSt                         
&& 0.184( 03)
&& 0.186( 03)
&& 0.190( 04)
&& 0.191( 03)
&& 0.171( 03)
&& 0.172( 03)
  &\cr\ifspace\myskip&&
&& 
&& 
&& 
&& 
&& 
&& 
  &\cr\fi\myskip\tlr
\myskip
&&  ChU_1                         
&& 0.174( 03)
&& 0.176( 03)
&& 0.182( 04)
&& 0.182( 03)
&& 0.162( 04)
&& 0.164( 03)
  &\cr\ifspace\myskip&&
&& 
&& 
&& 
&& 
&& 
&& 
  &\cr\fi\myskip\tlr
\myskip
&&  ChU_2                         
&& 0.170( 04)
&& 0.172( 04)
&& 0.177( 05)
&& 0.178( 04)
&& 0.157( 05)
&& 0.159( 03)
  &\cr\ifspace\myskip&&
&& 
&& 
&& 
&& 
&& 
&& 
  &\cr\fi\myskip\tlr
\myskip
&&  ChU_3                         
&& 0.166( 04)
&& 0.170( 05)
&& 0.175( 07)
&& 0.176( 05)
&& 0.154( 06)
&& 0.157( 04)
  &\cr\ifspace\myskip&&
&& 
&& 
&& 
&& 
&& 
&& 
  &\cr\fi\myskip\tlr
\myskip
&&  StSt                         
&& 0.291( 04)
&& 0.293( 05)
&& 0.303( 06)
&& 0.308( 05)
&& 0.268( 05)
&& 0.274( 04)
  &\cr\ifspace\myskip&&
&& 
&& 
&& 
&& 
&& 
&& 
  &\cr\fi\myskip\tlr
\myskip
&&  StU_1                         
&& 0.300( 04)
&& 0.301( 06)
&& 0.312( 07)
&& 0.318( 06)
&& 0.273( 06)
&& 0.282( 04)
  &\cr\ifspace\myskip&&
&& 
&& 
&& 
&& 
&& 
&& 
  &\cr\fi\myskip\tlr
\myskip
&&  StU_2                         
&& 0.302( 04)
&& 0.299( 09)
&& 0.314( 08)
&& 0.319( 06)
&& 0.273( 07)
&& 0.282( 05)
  &\cr\ifspace\myskip&&
&& 
&& 
&& 
&& 
&& 
&& 
  &\cr\fi\myskip\tlr
\myskip
&&  StU_3                         
&& 0.303( 05)
&& 0.297( 10)
&& 0.313( 11)
&& 0.318( 08)
&& 0.271( 09)
&& 0.281( 06)
  &\cr\ifspace\myskip&&
&& 
&& 
&& 
&& 
&& 
&& 
  &\cr\fi\myskip\tlr
\myskip
&&  U_1U_1                         
&& 0.316( 05)
&& 0.314( 05)
&& 0.329( 10)
&& 0.334( 05)
&& 0.285( 09)
&& 0.296( 04)
  &\cr\ifspace\myskip&&
&& 
&& 
&& 
&& 
&& 
&& 
  &\cr\fi\myskip\tlr
\myskip
&&  U_1U_2                         
&& 0.320( 05)
&& 0.317( 06)
&& 0.334( 13)
&& 0.338( 06)
&& 0.288( 11)
&& 0.299( 05)
  &\cr\ifspace\myskip&&
&& 
&& 
&& 
&& 
&& 
&& 
  &\cr\fi\myskip\tlr
\myskip
&&  U_1U_3                         
&& 0.322( 06)
&& 0.317( 06)
&& 0.333( 16)
&& 0.334( 09)
&& 0.288( 14)
&& 0.300( 05)
  &\cr\ifspace\myskip&&
&& 
&& 
&& 
&& 
&& 
&& 
  &\cr\fi\myskip\tlr
\myskip
&&  U_2U_2                         
&& 0.325( 06)
&& 0.320( 07)
&& 0.338( 17)
&& 0.337( 10)
&& 0.292( 15)
&& 0.303( 06)
  &\cr\ifspace\myskip&&
&& 
&& 
&& 
&& 
&& 
&& 
  &\cr\fi\myskip\tlr
\myskip
&&  U_2U_3                         
&& 0.326( 07)
&& 0.319( 08)
&& 0.334( 23)
&& 0.334( 13)
&& 0.292( 20)
&& 0.303( 07)
  &\cr\ifspace\myskip&&
&& 
&& 
&& 
&& 
&& 
&& 
  &\cr\fi\myskip\tlr
\myskip
&&  U_3U_3                         
&& 0.326( 07)
&& 0.316( 10)
&& 0.326( 31)
&& 0.327( 16)
&& 0.291( 27)
&& 0.302( 08)
  &\cr\ifspace\myskip&&
&& 
&& 
&& 
&& 
&& 
&& 
  &\cr\fi\myskip\tlr

\cr}}}}$$
 

%% file: t_frhoschemes.tex
$$
\let\ifspace=\iffalse
\def\myskip{\omit&height1.5pt&%
\omit&&%
\omit&&%
\omit&&%
\omit&&%
\omit&&%
\omit&&%
\omit&&%
 &\cr}
\vbox{\hbox{\vbox{
\tabskip=0pt\offinterlineskip
\def\tlr{\noalign{\hrule}}

\halign {\strut#& \vrule\vrule#\tabskip=3pt&
  \hfil$#$\hfil&\vrule#&
  \hfil$#$\hfil&\vrule#&
  \hfil$#$\hfil&\vrule#&
  \hfil$#$\hfil&\vrule#&
  \hfil$#$\hfil&\vrule#&
  \hfil$#$\hfil&\vrule#&
  \hfil$#$\hfil&\vrule#&
  \hfil$#$\hfil&\vrule#\tabskip=0pt\cr\tlr
\omit&height1.5pt&\multispan{15   }&\cr
\myskip
&& 
&& Z_{TADa}                      
&& Z_{TAD1}                      
&& Z_{TAD2}                      
&& Z_{TAD\pi}                    
&& Z_{TADU_0}                     
&& Z_{Tfg11}                     
&& Z_{BST\pi}                    
  &\cr
\myskip\tlr
\omit&height0.5pt&\multispan{15   }&\cr\tlr
\myskip
&&  ChCh                         
&& 0.184(2)
&& 0.186(2)
&& 0.193(2)
&& 0.196(3)
&& 0.173(2)
&& 0.188(2)
&& 0.119(2)
  &\cr\ifspace\myskip&&
&& 
&& 
&& 
&& 
&& 
&& 
&& 
  &\cr\fi\myskip\tlr
\myskip
&&  ChSt                         
&& 0.183(3)
&& 0.185(3)
&& 0.192(3)
&& 0.195(3)
&& 0.172(3)
&& 0.187(3)
&& 0.144(2)
  &\cr\ifspace\myskip&&
&& 
&& 
&& 
&& 
&& 
&& 
&& 
  &\cr\fi\myskip\tlr
\myskip
&&  ChU_1                         
&& 0.173(3)
&& 0.175(3)
&& 0.182(3)
&& 0.185(3)
&& 0.163(3)
&& 0.177(3)
&& 0.140(2)
  &\cr\ifspace\myskip&&
&& 
&& 
&& 
&& 
&& 
&& 
&& 
  &\cr\fi\myskip\tlr
\myskip
&&  ChU_2                         
&& 0.169(4)
&& 0.171(4)
&& 0.177(4)
&& 0.180(4)
&& 0.158(3)
&& 0.172(4)
&& 0.138(3)
  &\cr\ifspace\myskip&&
&& 
&& 
&& 
&& 
&& 
&& 
&& 
  &\cr\fi\myskip\tlr
\myskip
&&  ChU_3                         
&& 0.166(4)
&& 0.168(5)
&& 0.174(5)
&& 0.177(5)
&& 0.156(4)
&& 0.169(5)
&& 0.136(4)
  &\cr\ifspace\myskip&&
&& 
&& 
&& 
&& 
&& 
&& 
&& 
  &\cr\fi\myskip\tlr
\myskip
&&  StSt                         
&& 0.289(4)
&& 0.292(4)
&& 0.303(4)
&& 0.308(4)
&& 0.271(4)
&& 0.295(4)
&& 0.278(4)
  &\cr\ifspace\myskip&&
&& 
&& 
&& 
&& 
&& 
&& 
&& 
  &\cr\fi\myskip\tlr
\myskip
&&  StU_1                         
&& 0.297(5)
&& 0.301(5)
&& 0.312(5)
&& 0.317(5)
&& 0.279(4)
&& 0.304(5)
&& 0.294(5)
  &\cr\ifspace\myskip&&
&& 
&& 
&& 
&& 
&& 
&& 
&& 
  &\cr\fi\myskip\tlr
\myskip
&&  StU_2                         
&& 0.298(6)
&& 0.301(6)
&& 0.312(6)
&& 0.317(6)
&& 0.280(6)
&& 0.304(6)
&& 0.297(6)
  &\cr\ifspace\myskip&&
&& 
&& 
&& 
&& 
&& 
&& 
&& 
  &\cr\fi\myskip\tlr
\myskip
&&  StU_3                         
&& 0.297(7)
&& 0.300(7)
&& 0.311(7)
&& 0.317(7)
&& 0.279(6)
&& 0.303(7)
&& 0.298(7)
  &\cr\ifspace\myskip&&
&& 
&& 
&& 
&& 
&& 
&& 
&& 
  &\cr\fi\myskip\tlr
\myskip
&&  U_1U_1                         
&& 0.312(5)
&& 0.315(5)
&& 0.327(5)
&& 0.333(5)
&& 0.293(4)
&& 0.318(5)
&& 0.316(5)
  &\cr\ifspace\myskip&&
&& 
&& 
&& 
&& 
&& 
&& 
&& 
  &\cr\fi\myskip\tlr
\myskip
&&  U_1U_2                         
&& 0.315(5)
&& 0.319(5)
&& 0.331(5)
&& 0.336(5)
&& 0.296(5)
&& 0.322(5)
&& 0.322(5)
  &\cr\ifspace\myskip&&
&& 
&& 
&& 
&& 
&& 
&& 
&& 
  &\cr\fi\myskip\tlr
\myskip
&&  U_1U_3                         
&& 0.316(6)
&& 0.319(6)
&& 0.331(6)
&& 0.337(6)
&& 0.297(5)
&& 0.322(6)
&& 0.325(6)
  &\cr\ifspace\myskip&&
&& 
&& 
&& 
&& 
&& 
&& 
&& 
  &\cr\fi\myskip\tlr
\myskip
&&  U_2U_2                         
&& 0.319(6)
&& 0.322(6)
&& 0.335(6)
&& 0.340(6)
&& 0.300(6)
&& 0.325(6)
&& 0.329(6)
  &\cr\ifspace\myskip&&
&& 
&& 
&& 
&& 
&& 
&& 
&& 
  &\cr\fi\myskip\tlr
\myskip
&&  U_2U_3                         
&& 0.319(7)
&& 0.322(7)
&& 0.335(7)
&& 0.340(7)
&& 0.299(6)
&& 0.325(7)
&& 0.331(7)
  &\cr\ifspace\myskip&&
&& 
&& 
&& 
&& 
&& 
&& 
&& 
  &\cr\fi\myskip\tlr
\myskip
&&  U_3U_3                         
&& 0.318(8)
&& 0.321(8)
&& 0.334(8)
&& 0.339(8)
&& 0.299(7)
&& 0.324(8)
&& 0.332(8)
  &\cr\ifspace\myskip&&
&& 
&& 
&& 
&& 
&& 
&& 
&& 
  &\cr\fi\myskip\tlr
\cr}}}}$$
 

%% file: t_Zschemes.tex
\let\ifspace=\iffalse
\def\myskip{\omit&height1.5pt&%
\omit&&%
\omit&&%
\omit&&%
\omit&&%
\omit&&%
\omit&&%
\omit&&%
 &\cr}
\vbox{\hbox{\vbox{
\tabskip=0pt\offinterlineskip
\def\tlr{\noalign{\hrule}}
\halign {\strut#& \vrule\vrule#\tabskip=3pt&
  \hfil$#$\hfil&\vrule#&
  \hfil$#$\hfil&\vrule#&
  \hfil$#$\hfil&\vrule#&
  \hfil$#$\hfil&\vrule#&
  \hfil$#$\hfil&\vrule#&
  \hfil$#$\hfil&\vrule#&
  \hfil$#$\hfil&\vrule#&
  \hfil$#$\hfil&\vrule#\tabskip=0pt\cr\tlr
\omit&height1.5pt&\multispan{15   }&\cr
\myskip
&& 
&& Z_{TADa}                    
&& Z_{TAD1}                    
&& Z_{TAD2}                    
&& Z_{TAD\pi}                   
&& Z_{TADU_0}                   
&& Z_{Tgf11}                   
&& Z_{Boost\pi}                   
  &\cr
\myskip\tlr
\omit&height0.5pt&\multispan{15   }&\cr\tlr
\myskip
&& Z_\psi                      
&& \displaystyle 1 - {3\kappa \over 4\kappa_c}
&& \displaystyle 1 - {3\kappa \over 4\kappa_c}
&& \displaystyle 1 - {3\kappa \over 4\kappa_c}
&& \displaystyle 1 - {3\kappa \over 4\kappa_c}
&& \displaystyle 1 - {3\kappa \over 4\kappa_c}
&& \displaystyle 1 - {3\kappa \over 4\kappa_c}
&& \displaystyle 2\kappa
  &\cr\ifspace\myskip&&
&& 
&& 
&& 
&& 
&& 
&& 
&& 
  &\cr\fi\myskip\tlr
\myskip
&& {\rm Tadpole}                      
&& 1/8\kappa_c
&& 1/8\kappa_c
&& 1/8\kappa_c
&& 1/8\kappa_c
&& U_0
&& 1/8\kappa_c
&& {\rm NO}
  &\cr\ifspace\myskip&&
&& 
&& 
&& 
&& 
&& 
&& 
&& 
  &\cr\fi\myskip\tlr
\myskip
&& q^*                     
&& 2\ GeV
&& 1/a
&& 2/a
&& \pi/a
&& 1/a
&& 1/a
&& \pi/a
  &\cr\ifspace\myskip&&
&& 
&& 
&& 
&& 
&& 
&& 
&& 
  &\cr\fi\myskip\tlr
\myskip
&& \alpha_s(q^*)
&& 0.202
&& 0.190
&& 0.151
&& 0.133
&& 0.190
&& 0.181
&& 0.133
  &\cr\ifspace\myskip&&
&& 
&& 
&& 
&& 
&& 
&& 
&& 
  &\cr\fi\myskip\tlr
\cr}}}}

%% file: t_mstab.tex
\let\ifspace=\iffalse
\def\myskip{\omit&height1.5pt&%
\omit&&%
\omit&&%
\omit&&%
 &\cr}
\vbox{\hbox{\vbox{
\tabskip=0pt\offinterlineskip
\def\tlr{\noalign{\hrule}}
\halign {\strut#& \vrule\vrule#\tabskip=3pt&
  \hfil$#$\hfil&\vrule#&
  \hfil$#$\hfil&\vrule#&
  \hfil$#$\hfil&\vrule#&
  \hfil$#$\hfil&\vrule#\tabskip=0pt\cr\tlr
\omit&height1.5pt&\multispan{7   }&\cr
\myskip
&& 
&& \kappa_s
&& m_{s,np} a                    
&& m_s\ (2\ GeV)\ MeV
  &\cr
\myskip\tlr
\omit&height0.5pt&\multispan{7   }&\cr\tlr
\myskip
&& M_K^2 / M_\rho^2
&& 0.15503(7)
&& 0.0372(14)
&& 129(2)
  &\cr\ifspace\myskip&&
&& 
&& 
&& 
  &\cr\fi\myskip\tlr
\myskip
&& M_{K^*} / M_\rho
&& 0.15479(19)
&& 0.0421(36)
&& 145(9)
  &\cr\ifspace\myskip&&
&& 
&& 
&& 
  &\cr\fi\myskip\tlr
\myskip
&& M_\phi / M_\rho                     
&& 0.15464(17)
&& 0.0445(32)
&& 154(8)
  &\cr\ifspace\myskip&&
&& 
&& 
&& 
  &\cr\fi\myskip\tlr
\cr}}}}

%% file: t_Dmass.tex
\let\ifspace=\iffalse
\def\myskip{\omit&height1.5pt&%
\omit&&%
\omit&&%
\omit&&%
 &\cr}
\vbox{\hbox{\vbox{
\tabskip=0pt\offinterlineskip
\def\tlr{\noalign{\hrule}}
\halign {\strut#& \vrule\vrule#\tabskip=3pt&
  \hfil$#$\hfil&\vrule#&
  \hfil$#$\hfil&\vrule#&
  \hfil$#$\hfil&\vrule#&
  \hfil$#$\hfil&\vrule#\tabskip=0pt\cr\tlr
\omit&height1.5pt&\multispan{7   }&\cr
\myskip
&& 
&& M_1
&& M_2
&& Expt.
  &\cr
\myskip\tlr
\omit&height0.5pt&\multispan{7   }&\cr\tlr
\myskip
&& M_D
&& 1805(31)
&& 1990(34)
&& 1869
  &\cr\ifspace\myskip&&
&& 
&& 
&& 
  &\cr\fi\myskip\tlr
\myskip
&& M_{D^*}
&& 1876(32)
&& 2085(35)
&& 2008
  &\cr\ifspace\myskip&&
&& 
&& 
&& 
  &\cr\fi\myskip\tlr
\myskip
&& M_{D_s}(m_s(M_K))
&& 1896(30)
&& 2112(32)
&& 1969
  &\cr\ifspace\myskip&&
&& 
&& 
&& 
  &\cr\fi\myskip\tlr
\myskip
&& M_{D_s}(m_s(M_\phi))
&& 1914(26)
&& 2137(27)
&& 1969
  &\cr\ifspace\myskip&&
&& 
&& 
&& 
  &\cr\fi\myskip\tlr
\myskip
&& M_{D_s^*}(m_s(M_K))
&& 1961(31)
&& 2201(34)
&& 2110?
  &\cr\ifspace\myskip&&
&& 
&& 
&& 
  &\cr\fi\myskip\tlr
\myskip
&& M_{D_s^*}(m_s(M_\phi))
&& 1978(27)
&& 2224(29)
&& 2110?
  &\cr\ifspace\myskip&&
&& 
&& 
&& 
  &\cr\fi\myskip\tlr
\cr}}}}

%% file: t_fpilat.tex
$$
\let\ifspace=\iffalse
\def\myskip{\omit&height1.5pt&%
\omit&&%
\omit&&%
\omit&&%
\omit&&%
\omit&&%
\omit&&%
\omit&&%
 &\cr}
\vbox{\hbox{\vbox{
\tabskip=0pt\offinterlineskip
\def\tlr{\noalign{\hrule}}

\halign {\strut#& \vrule\vrule#\tabskip=0.5pt&
  \hfil$#$\hfil&\vrule#&
  \hfil$#$\hfil&\vrule#&
  \hfil$#$\hfil&\vrule#&
  \hfil$#$\hfil&\vrule#&
  \hfil$#$\hfil&\vrule#&
  \hfil$#$\hfil&\vrule#&
  \hfil$#$\hfil&\vrule#&
  \hfil$#$\hfil&\vrule#\tabskip=0pt\cr\tlr
\omit&height1.5pt&\multispan{15   }&\cr
\myskip
&& 
&& Z_{TADa}                      
&& Z_{TAD1}                      
&& Z_{TAD2}                      
&& Z_{TAD\pi}                    
&& Z_{TADU_0}                     
&& Z_{Tgf11}                     
&& Z_{Boost\pi}                  
  &\cr
\myskip\tlr
\omit&height0.5pt&\multispan{15   }&\cr\tlr
\myskip
&& (M_1)$\hfill$\quad f_\pi                                    
&& 0.057( 01)
&& 0.058( 01)
&& 0.058( 01)
&& 0.059( 01)
&& 0.054( 01)
&& 0.058( 01)
&& 0.060( 01)
  &\cr\ifspace\myskip&&
&& 
&& 
&& 
&& 
&& 
&& 
&& 
  &\cr\fi\myskip\tlr
\myskip
&& (M_1,M_K)$\hfill$\quad f_K                                  
&& 0.067( 01)
&& 0.067( 01)
&& 0.068( 01)
&& 0.068( 01)
&& 0.063( 01)
&& 0.067( 01)
&& 0.068( 01)
  &\cr\ifspace\myskip&&
&& 
&& 
&& 
&& 
&& 
&& 
&& 
  &\cr\fi\myskip\tlr
\myskip
&& (M_1,M_\phi)$\hfill$\quad f_K                               
&& 0.068( 01)
&& 0.068( 01)
&& 0.069( 01)
&& 0.070( 01)
&& 0.064( 01)
&& 0.069( 01)
&& 0.069( 01)
  &\cr\ifspace\myskip&&
&& 
&& 
&& 
&& 
&& 
&& 
&& 
  &\cr\fi\myskip\tlr
\myskip
&& (M_1,M_K)$\hfill$\quad f_K/f_\pi                            
&& 1.161( 11)
&& 1.161( 11)
&& 1.161( 11)
&& 1.161( 11)
&& 1.161( 11)
&& 1.161( 11)
&& 1.127( 10)
  &\cr\ifspace\myskip&&
&& 
&& 
&& 
&& 
&& 
&& 
&& 
  &\cr\fi\myskip\tlr
\myskip
&& (M_1,M_\phi)$\hfill$\quad f_K/f_\pi                         
&& 1.186( 16)
&& 1.186( 16)
&& 1.186( 16)
&& 1.186( 16)
&& 1.186( 16)
&& 1.186( 16)
&& 1.145( 14)
  &\cr\ifspace\myskip&&
&& 
&& 
&& 
&& 
&& 
&& 
&& 
  &\cr\fi\myskip\tlr
\myskip
&& (M_1)$\hfill$\quad f_D                                      
&& 0.103( 03)
&& 0.103( 03)
&& 0.105( 03)
&& 0.105( 03)
&& 0.097( 03)
&& 0.104( 03)
&& 0.083( 02)
  &\cr\ifspace\myskip&&
&& 
&& 
&& 
&& 
&& 
&& 
&& 
  &\cr\fi\myskip\tlr
\myskip
&& (M_2)$\hfill$\quad f_D                                      
&& 0.098( 02)
&& 0.098( 03)
&& 0.100( 03)
&& 0.100( 03)
&& 0.092( 02)
&& 0.099( 03)
&& 0.079( 02)
  &\cr\ifspace\myskip&&
&& 
&& 
&& 
&& 
&& 
&& 
&& 
  &\cr\fi\myskip\tlr
\myskip
&& (M_1)$\hfill$\quad f_D/f_\pi                                
&& 1.793( 49)
&& 1.793( 49)
&& 1.793( 49)
&& 1.793( 49)
&& 1.793( 49)
&& 1.793( 49)
&& 1.388( 38)
  &\cr\ifspace\myskip&&
&& 
&& 
&& 
&& 
&& 
&& 
&& 
  &\cr\fi\myskip\tlr
\myskip
&& (M_2)$\hfill$\quad f_D/f_\pi                                
&& 1.705( 45)
&& 1.705( 45)
&& 1.705( 45)
&& 1.705( 45)
&& 1.705( 45)
&& 1.705( 45)
&& 1.320( 35)
  &\cr\ifspace\myskip&&
&& 
&& 
&& 
&& 
&& 
&& 
&& 
  &\cr\fi\myskip\tlr
\myskip
&& (M_1,M_K)$\hfill$\quad f_{D_s}                              
&& 0.115( 02)
&& 0.115( 02)
&& 0.117( 02)
&& 0.118( 02)
&& 0.108( 02)
&& 0.116( 02)
&& 0.091( 01)
  &\cr\ifspace\myskip&&
&& 
&& 
&& 
&& 
&& 
&& 
&& 
  &\cr\fi\myskip\tlr
\myskip
&& (M_2,M_K)$\hfill$\quad f_{D_s}                              
&& 0.109( 02)
&& 0.109( 02)
&& 0.111( 02)
&& 0.111( 02)
&& 0.102( 02)
&& 0.110( 02)
&& 0.086( 01)
  &\cr\ifspace\myskip&&
&& 
&& 
&& 
&& 
&& 
&& 
&& 
  &\cr\fi\myskip\tlr
\myskip
&& (M_1,M_\phi)$\hfill$\quad f_{D_s}                           
&& 0.118( 02)
&& 0.118( 02)
&& 0.120( 02)
&& 0.120( 02)
&& 0.110( 02)
&& 0.118( 02)
&& 0.092( 01)
  &\cr\ifspace\myskip&&
&& 
&& 
&& 
&& 
&& 
&& 
&& 
  &\cr\fi\myskip\tlr
\myskip
&& (M_2,M_\phi)$\hfill$\quad f_{D_s}                           
&& 0.111( 02)
&& 0.112( 02)
&& 0.113( 02)
&& 0.114( 02)
&& 0.104( 02)
&& 0.112( 02)
&& 0.087( 01)
  &\cr\ifspace\myskip&&
&& 
&& 
&& 
&& 
&& 
&& 
&& 
  &\cr\fi\myskip\tlr
\myskip
&& (M_1,M_K)$\hfill$\quad f_{D_s}/f_D                          
&& 1.117( 19)
&& 1.117( 19)
&& 1.117( 19)
&& 1.117( 19)
&& 1.117( 19)
&& 1.117( 19)
&& 1.088( 18)
  &\cr\ifspace\myskip&&
&& 
&& 
&& 
&& 
&& 
&& 
&& 
  &\cr\fi\myskip\tlr
\myskip
&& (M_2,M_K)$\hfill$\quad f_{D_s}/f_D                          
&& 1.112( 18)
&& 1.112( 18)
&& 1.112( 18)
&& 1.112( 18)
&& 1.112( 18)
&& 1.112( 18)
&& 1.083( 17)
  &\cr\ifspace\myskip&&
&& 
&& 
&& 
&& 
&& 
&& 
&& 
  &\cr\fi\myskip\tlr
\myskip
&& (M_1,M_\phi)$\hfill$\quad f_{D_s}/f_D                       
&& 1.141( 22)
&& 1.141( 22)
&& 1.141( 22)
&& 1.141( 22)
&& 1.141( 22)
&& 1.141( 22)
&& 1.106( 20)
  &\cr\ifspace\myskip&&
&& 
&& 
&& 
&& 
&& 
&& 
&& 
  &\cr\fi\myskip\tlr
\myskip
&& (M_2,M_\phi)$\hfill$\quad f_{D_s}/f_D                       
&& 1.135( 21)
&& 1.135( 21)
&& 1.135( 21)
&& 1.135( 21)
&& 1.135( 21)
&& 1.135( 21)
&& 1.100( 19)
  &\cr\ifspace\myskip&&
&& 
&& 
&& 
&& 
&& 
&& 
&& 
  &\cr\fi\myskip\tlr

\cr}}}}$$
 

%% file: t_dcTAD.tex
$$
\let\ifspace=\iffalse
\def\myskip{\omit&height1.5pt&%
\omit&&%
\omit&&%
\omit&&%
\omit&&%
 &\cr}
\vbox{\hbox{\vbox{
\tabskip=0pt\offinterlineskip
\def\tlr{\noalign{\hrule}}

\halign {\strut#& \vrule\vrule#\tabskip=3pt&
  \hfil$#$\hfil&\vrule#&
  \hfil$#$\hfil&\vrule#&
  \hfil$#$\hfil&\vrule#&
  \hfil$#$\hfil&\vrule#&
  \hfil$#$\hfil&\vrule#\tabskip=0pt\cr\tlr
\omit&height1.5pt&\multispan{ 9   }&\cr
\myskip
&& 
&& M_1\ \&\ m_s(m_K)             
&& M_1\ \&\ m_s(m_\phi)          
&& M_2\ \&\ m_s(m_K)             
&& M_2\ \&\ m_s(m_\phi)          
  &\cr
\myskip\tlr
\omit&height0.5pt&\multispan{ 9   }&\cr\tlr
\myskip
&& f_\pi                         
&& 134.4( 41)
&& 
&& 
&& 
  &\cr\ifspace\myskip&&
&& 
&& 
&& 
&& 
  &\cr\fi\myskip\tlr
\myskip
&& f_K                           
&& 156.1( 37)
&& 159.4( 33)
&& 
&& 
  &\cr\ifspace\myskip&&
&& 
&& 
&& 
&& 
  &\cr\fi\myskip\tlr
\myskip
&& f_D                           
&& 241.0( 75)
&& 
&& 229.2( 70)
&& 
  &\cr\ifspace\myskip&&
&& 
&& 
&& 
&& 
  &\cr\fi\myskip\tlr
\myskip
&& f_{D_s}                       
&& 269.1( 54)
&& 275.0( 46)
&& 254.8( 51)
&& 260.1( 44)
  &\cr\ifspace\myskip&&
&& 
&& 
&& 
&& 
  &\cr\fi\myskip\tlr
\cr}}}}$$
 

%% file: t_dcfinal.tex
\let\ifspace=\iffalse
\def\myskip{\omit&height1.5pt&%
\omit&&%
\omit&&%
\omit&&%
\omit&&%
\omit&&%
\omit&&%
\omit&&%
 &\cr}
\vbox{\hbox{\vbox{
\tabskip=0pt\offinterlineskip
\def\tlr{\noalign{\hrule}}
\halign {\strut#& \vrule\vrule#\tabskip=3pt&
  \hfil$#$\hfil&\vrule#&
  \hfil$#$\hfil&\vrule#&
  \hfil$#$\hfil&\vrule#&
  \hfil$#$\hfil&\vrule#&
  \hfil$#$\hfil&\vrule#&
  \hfil$#$\hfil&\vrule#&
  \hfil$#$\hfil&\vrule#&
  \hfil$#$\hfil&\vrule#\tabskip=0pt\cr\tlr
\omit&height1.5pt&\multispan{15   }&\cr
\myskip
&& 
&& {\rm Best }
&& {\rm Statistical\ \&}
&& {\rm Tuning }                
&& {\rm Tuning }                  
&&                    
&& {\rm Tuning}
&&                    
  &\cr
\myskip
&& 
&& {\rm Estimate }
&& {\rm Extrapolation}
&& m_s          
&& m_c
&& q^*
&& a\ (3\%)
&& Z_A                
  &\cr
\myskip\tlr
\omit&height0.5pt&\multispan{15   }&\cr\tlr
\myskip
&& f_\pi           
&& 134
&& 4
&& -
&& -
&& +2
&& 4
&& 10
  &\cr\ifspace\myskip&&
&& 
&& 
&& 
&&
&& 
&& 
&& 
&& 
  &\cr\fi\myskip\tlr
\myskip
&& f_K
&& 159 
&& 3
&& -3 
&& -
&& +3
&& 5
&& 10
  &\cr\ifspace\myskip&&
&& 
&& 
&& 
&&
&& 
&& 
&& 
&& 
  &\cr\fi\myskip\tlr
\myskip
&& f_D
&& 229
&& 7
&& -
&& +12
&& +4
&& 7
&& 14
  &\cr\ifspace\myskip&&
&& 
&& 
&& 
&& 
&& 
&& 
&& 
&& 
  &\cr\fi\myskip\tlr
\myskip
&& f_{D_s}
&& 260
&& 4 
&& -5
&& +15
&& +4
&& 8
&& 20
  &\cr\ifspace\myskip&&
&& 
&& 
&& 
&& 
&& 
&& 
&& 
&& 
  &\cr\fi\myskip\tlr
\myskip
&& f_K/f_\pi
&& 1.19
&& 0.02
&& -0.025
&& -
&& -
&& -
&& 0
  &\cr\ifspace\myskip&&
&& 
&& 
&& 
&& 
&& 
&& 
&& 
&& 
  &\cr\fi\myskip\tlr
\myskip
&& f_D/f_\pi
&& 1.71
&& 0.05
&& -
&& +0.09
&& -
&& -
&& ?
  &\cr\ifspace\myskip&&
&& 
&& 
&& 
&& 
&& 
&& 
&& 
&& 
  &\cr\fi\myskip\tlr
\myskip
&& f_{D_s}/f_D
&& 1.135
&& 0.021
&& -0.023
&& +0.006
&& -
&& -
&& 0
  &\cr\ifspace\myskip&&
&& 
&& 
&& 
&& 
&& 
&& 
&& 
&& 
  &\cr\fi\myskip\tlr
\cr}}}}

%% file: t_fVlat.tex
$$
\let\ifspace=\iffalse
\def\myskip{\omit&height1.5pt&%
\omit&&%
\omit&&%
\omit&&%
\omit&&%
\omit&&%
\omit&&%
\omit&&%
 &\cr}
\vbox{\hbox{\vbox{
\tabskip=0pt\offinterlineskip
\def\tlr{\noalign{\hrule}}

\halign {\strut#& \vrule\vrule#\tabskip=1pt&
  \hfil$#$\hfil&\vrule#&
  \hfil$#$\hfil&\vrule#&
  \hfil$#$\hfil&\vrule#&
  \hfil$#$\hfil&\vrule#&
  \hfil$#$\hfil&\vrule#&
  \hfil$#$\hfil&\vrule#&
  \hfil$#$\hfil&\vrule#&
  \hfil$#$\hfil&\vrule#\tabskip=0pt\cr\tlr
\omit&height1.5pt&\multispan{15   }&\cr
\myskip
&& 
&& Z_{TADa}                      
&& Z_{TAD1}                      
&& Z_{TAD2}                      
&& Z_{TAD\pi}                    
&& Z_{TAD8k}                     
&& Z_{Tgf11}                     
&& Z_{Boost\pi}                  
  &\cr
\myskip\tlr
\omit&height0.5pt&\multispan{15   }&\cr\tlr
\myskip
&& (M_1)$\hfill$\  [M_\rho]
&& 0.324( 10)
&& 0.328( 10)
&& 0.340( 11)
&& 0.346( 11)
&& 0.305( 09)
&& 0.331( 10)
&& 0.345( 11)
  &\cr\ifspace\myskip&&
&& 
&& 
&& 
&& 
&& 
&& 
&& 
  &\cr\fi\myskip\tlr
\myskip
&& (M_1,M_K)$\hfill$\  [M_{K^*}]
&& 0.319( 06)
&& 0.322( 06)
&& 0.335( 07)
&& 0.340( 07)
&& 0.300( 06)
&& 0.325( 06)
&& 0.331( 06)
  &\cr\ifspace\myskip&&
&& 
&& 
&& 
&& 
&& 
&& 
&& 
  &\cr\fi\myskip\tlr
\myskip
&& (M_1,M_\phi)$\hfill$\  [M_{K^*}]
&& 0.315( 06)
&& 0.318( 06)
&& 0.330( 06)
&& 0.336( 06)
&& 0.296( 06)
&& 0.321( 06)
&& 0.325( 06)
  &\cr\ifspace\myskip&&
&& 
&& 
&& 
&& 
&& 
&& 
&& 
  &\cr\fi\myskip\tlr
\myskip
&& (M_1,M_K)$\hfill$\  [M_{\phi}]
&& 0.312( 04)
&& 0.316( 04)
&& 0.328( 05)
&& 0.333( 05)
&& 0.293( 04)
&& 0.318( 04)
&& 0.316( 04)
  &\cr\ifspace\myskip&&
&& 
&& 
&& 
&& 
&& 
&& 
&& 
  &\cr\fi\myskip\tlr
\myskip
&& (M_1,M_\phi)$\hfill$\  [M_{\phi}]
&& 0.308( 04)
&& 0.311( 04)
&& 0.323( 05)
&& 0.328( 05)
&& 0.289( 04)
&& 0.314( 04)
&& 0.309( 05)
  &\cr\ifspace\myskip&&
&& 
&& 
&& 
&& 
&& 
&& 
&& 
  &\cr\fi\myskip\tlr
\myskip
&& (M_1)$\hfill$\  [M_{D^*}]
&& 0.162( 05)
&& 0.164( 05)
&& 0.170( 06)
&& 0.173( 06)
&& 0.152( 05)
&& 0.165( 06)
&& 0.134( 04)
  &\cr\ifspace\myskip&&
&& 
&& 
&& 
&& 
&& 
&& 
&& 
  &\cr\fi\myskip\tlr
\myskip
&& (M_2)$\hfill$\  [M_{D^*}]
&& 0.137( 04)
&& 0.139( 04)
&& 0.144( 04)
&& 0.147( 04)
&& 0.129( 04)
&& 0.140( 04)
&& 0.114( 03)
  &\cr\ifspace\myskip&&
&& 
&& 
&& 
&& 
&& 
&& 
&& 
  &\cr\fi\myskip\tlr
\myskip
&& (M_1,M_K)$\hfill$\  [M_{D_s^*}]
&& 0.173( 03)
&& 0.175( 03)
&& 0.182( 03)
&& 0.185( 03)
&& 0.163( 03)
&& 0.177( 03)
&& 0.140( 03)
  &\cr\ifspace\myskip&&
&& 
&& 
&& 
&& 
&& 
&& 
&& 
  &\cr\fi\myskip\tlr
\myskip
&& (M_2,M_K)$\hfill$\  [M_{D_s^*}]
&& 0.145( 02)
&& 0.147( 02)
&& 0.152( 03)
&& 0.155( 03)
&& 0.136( 02)
&& 0.148( 02)
&& 0.117( 02)
  &\cr\ifspace\myskip&&
&& 
&& 
&& 
&& 
&& 
&& 
&& 
  &\cr\fi\myskip\tlr
\myskip
&& (M_1,M_\phi)$\hfill$\  [M_{D_s^*}]
&& 0.175( 03)
&& 0.177( 03)
&& 0.184( 03)
&& 0.187( 03)
&& 0.164( 03)
&& 0.179( 03)
&& 0.141( 03)
  &\cr\ifspace\myskip&&
&& 
&& 
&& 
&& 
&& 
&& 
&& 
  &\cr\fi\myskip\tlr
\myskip
&& (M_2,M_\phi)$\hfill$\  [M_{D_s^*}]
&& 0.146( 02)
&& 0.148( 02)
&& 0.154( 03)
&& 0.156( 03)
&& 0.138( 02)
&& 0.149( 02)
&& 0.118( 02)
  &\cr\ifspace\myskip&&
&& 
&& 
&& 
&& 
&& 
&& 
&& 
  &\cr\fi\myskip\tlr

\cr}}}}$$

%% file: t_fVcurrent.tex
$$
\let\ifspace=\iffalse
\def\myskip{\omit&height1.5pt&%
\omit&&%
\omit&&%
\omit&&%
 &\cr}
\vbox{\hbox{\vbox{
\tabskip=0pt\offinterlineskip
\def\tlr{\noalign{\hrule}}

\halign {\strut#& \vrule\vrule#\tabskip=3pt&
  \hfil$#$\hfil&\vrule#&
  \hfil$#$\hfil&\vrule#&
  \hfil$#$\hfil&\vrule#&
  \hfil$#$\hfil&\vrule#\tabskip=0pt\cr\tlr
\omit&height1.5pt&\multispan{ 7   }&\cr
\myskip
&& 
&& LOCAL                                                     
&& EXTENDED                                                  
&& CONSERVED                                                 
  &\cr
\myskip\tlr
\omit&height0.5pt&\multispan{ 7   }&\cr\tlr
\myskip
&& (M_1)$\hfill$\quad [M_\rho]
&& 0.328( 10)
&& 0.335( 26)
&& 0.304( 18)
  &\cr\ifspace\myskip&&
&& 
&& 
&& 
  &\cr\fi\myskip\tlr
\myskip
&& (M_1,M_K)$\hfill$\quad [M_{K^*}]
&& 0.322( 06)
&& 0.338( 13)
&& 0.297( 10)
  &\cr\ifspace\myskip&&
&& 
&& 
&& 
  &\cr\fi\myskip\tlr
\myskip
&& (M_1,M_\phi)$\hfill$\quad [M_{K^*}]
&& 0.318( 06)
&& 0.334( 12)
&& 0.293( 09)
  &\cr\ifspace\myskip&&
&& 
&& 
&& 
  &\cr\fi\myskip\tlr
\myskip
&& (M_1,M_K)$\hfill$\quad [M_{\phi}]
&& 0.316( 04)
&& 0.332( 06)
&& 0.291( 06)
  &\cr\ifspace\myskip&&
&& 
&& 
&& 
  &\cr\fi\myskip\tlr
\myskip
&& (M_1,M_\phi)$\hfill$\quad [M_{\phi}]
&& 0.311( 04)
&& 0.327( 06)
&& 0.287( 05)
  &\cr\ifspace\myskip&&
&& 
&& 
&& 
  &\cr\fi\myskip\tlr
\myskip
&& (M_1)$\hfill$\quad [M_{D^*}]
&& 0.164( 05)
&& 0.171( 06)
&& 0.151( 05)
  &\cr\ifspace\myskip&&
&& 
&& 
&& 
  &\cr\fi\myskip\tlr
\myskip
&& (M_2)$\hfill$\quad [M_{D^*}]
&& 0.139( 04)
&& 0.146( 05)
&& 0.128( 04)
  &\cr\ifspace\myskip&&
&& 
&& 
&& 
  &\cr\fi\myskip\tlr
\myskip
&& (M_1,M_K)$\hfill$\quad [M_{D_s^*}]
&& 0.175( 03)
&& 0.182( 03)
&& 0.163( 03)
  &\cr\ifspace\myskip&&
&& 
&& 
&& 
  &\cr\fi\myskip\tlr
\myskip
&& (M_2,M_K)$\hfill$\quad [M_{D_s^*}]
&& 0.147( 02)
&& 0.152( 03)
&& 0.137( 03)
  &\cr\ifspace\myskip&&
&& 
&& 
&& 
  &\cr\fi\myskip\tlr
\myskip
&& (M_1,M_\phi)$\hfill$\quad [M_{D_s^*}]
&& 0.177( 03)
&& 0.183( 03)
&& 0.164( 03)
  &\cr\ifspace\myskip&&
&& 
&& 
&& 
  &\cr\fi\myskip\tlr
\myskip
&& (M_2,M_\phi)$\hfill$\quad [M_{D_s^*}]
&& 0.148( 02)
&& 0.153( 03)
&& 0.138( 03)
  &\cr\ifspace\myskip&&
&& 
&& 
&& 
  &\cr\fi\myskip\tlr
\cr}}}}$$
 